\documentstyle[12pt]{article}
\makeatletter
\let\chapter\hid@chapter
\makeatother
\input psfig.tex
\def\lsim{\lower.5ex\hbox{$\; \buildrel < \over \sim \;$}}
\def\gsim{\lower.5ex\hbox{$\; \buildrel > \over \sim \;$}}
\begin{document}
\pagenumbering{arabic}
\title{Identification of Astrophysical Black Holes}

\author{Sandip K. Chakrabarti\\
S.N. Bose National Centre for Basic Sciences\\
JD-Block, Sector-III, Salt Lake, Calcutta 700091, INDIA}

\maketitle

\begin{abstract}

Black holes are by definition {\it black}, and therefore cannot be
directly observed by using electromagnetic radiations. Convincing 
identification of black holes must necessarily depend on the 
identification of a very specially behaving matter and radiation which 
surround them. A major problem in this subject of black hole astrophysics 
is to quantify the behaviour of matter and radiation close to the horizon.
In this review, the subject of black hole accretion and
outflow is systematically developed. It is shown that
both the stationary as well as the non-stationary properties 
of the observed spectra could be generally understood by these solutions.
It is suggested that the solutions of radiative hydrodynamic 
equations may produce clear spectral signatures of black holes. 
Other circumstantial evidences of black holes, both in the galactic 
centers as well as in binary systems, are also presented.

\end{abstract}

\noindent ACCEPTED FOR PUBLICATION (March 2nd, 1998) 
IN INDIAN JOURNAL OF PHYSICS (REVIEW SECTION).

\section{Introduction}

Stellar mass black holes are the end products of stars. After the 
fuel is exhausted inside a normal star, the core collapses and 
the supernova explosion occurs. If the mass of the core is lower 
than, say, $\sim 3M_\odot$, the object formed at the center may be a 
neutron star. Otherwise, it is a black hole. Therefore, some of the 
compact binary systems should contain black holes. Similarly, 
core collapse in the proto-galactic phase could also produce 
supermassive black holes ($M \sim 10^6$ to $10^9M_\odot$). In spiral 
galaxies, the central black holes are less massive (say, $10^{6-7}\  
M_\odot$), while in elliptical galaxies the central black holes are 
more massive (say, a few times $10^{8-9}\ M_\odot$). 

Astrophysical community generally believes that the black holes should 
exist because of the solid foundation of the theory of general 
relativity which predicts them. The problem remains that of identification. 
Black holes do not emit anything except Hawking radiation, which, for any 
typical mass of the astrophysical black holes is so cold 
(typically $60$ nano Kelvin for a solar mass black hole, and goes down 
inversely with increase in mass) that it would be 
entirely masked by the much hotter microwave background radiation.
Classically, black holes are point-like with infinite density and are 
surrounded by an imaginary one-way membrane called `event horizon'
of radius $R_g=2GM_{BH}/c^2$. Here, $G$ and $c$ are gravitational constant and 
velocity of light respectively, $M_{BH}$ is the mass of the black hole.
$R_g$ is known as the Schwarzschild radius and is roughly equal to 
$30$ kilometers for a $10\ M_\odot$ black hole. For a maximally rotating
(Kerr) black hole, the radius is half as small. Surrounding matter and 
radiation are pulled by the black hole only to disappear inside 
never to be observed again. Not even light, what to talk about matter, can escape 
to distant observers from regions within the horizon, making it {\it 
impossible} to detect a black hole through direct observations. A positive 
identification must therefore rely on indirect and circumstantial evidences. 
In fact, the problem of identification of black holes boils down to the 
identification of surrounding matter which may behave in a `funny' way. We 
shall quantify the degree of `funniness' as we go along.

In this {\it review}, we discuss how a black hole could be identified.
We first present elementary properties of the spacetime around a 
black hole and compare them with those of a Newtonian star. We discuss
in great length the properties of the global solutions of equations
which govern the behaviour of matter. We then show that the observations 
in the last couple of decades do agree with these properties. Towards the end 
we make a comparative study of methodologies of black hole detection 
and present our judgment on the best way to detect black holes. 

\section{Behaviour of Matter Around a Black Hole: Theoretical Expectations}

Generally speaking,  we shall use geometrical units where
masses, lengths and times are measured in units of mass of the
black hole $M_{BH}$, the Schwarzschild radius of the black hole
$r_g=2GM_{BH}/c^2$ and the light crossing time $t_g=2GM_{BH}/c^3$
of the black hole respectively. Since this is a review, and some figures
of other works with different conventions had to be borrowed, we may use $GM_{BH}/c^3$
for the length scale for the lengthscale, instead. Radial distances would be 
generally denoted by $r$ and in component form $x=r$ would be used.
Angular momentum would be denoted by $l$ for inviscid flows and $\lambda$ 
for viscous flows. We shall mention the choice of units whenever 
any confusion arises. Sometimes we may revert back to cgs unit 
when we need to put in numbers.

A few elementary definitions in this context are in order:  The
process by which matter falls into a black hole or a neutron
star (or, to any star, in general) is known as {\it accretion}. 
The accretion rate refers to the rate (in units of, e.g., 
$gms/sec$) at which matter falls into a black hole at a 
certain radius. For a steady state accretion, this rate is
fixed at all radii. Matter usually comes with some angular 
momentum. A Keplerian distribution of angular momentum is
achieved when the centrifugal force of the matter that spirals into
the black hole matches with the gravitational force acting on it. 
If other forces, such as that due to radiation pressure, ion pressure,
inertial force etc. are operating, the disk need not be Keplerian.

Followings are the estimates of physical quantities around black holes:

\indent 1. {\it Length scale}: $r_g$. Physical quantities are expected to
have variations in not too smaller that this length. Similarly,
if there are perturbations on accretion disks, the size of the
perturbations are also of similar length. $r_g \sim 3 \times 
10^5 \frac{M_{BH}}{M_\odot}$cm. In comparison, the sun has a 
radius of $r_g \sim 7 \times 10^{10}$cm, or, roughly two-tenths
of a million Schwarzschild radii.

\indent 2. {\it Time scale of variabilities}: $t_g=r_g/c$. This would be the 
shortest time scale of variation of quantities close to the black hole 
horizon. $t_v \sim \sim 3 \times 10^{-5} \frac{M}{M_\odot}$s.

\indent 3. {\it Specific angular momentum}: $l_g=r_g c$. Matter with this angular 
momentum has a centrifugal force comparable to the gravitational force. $l_g 
\sim 9 \times 10^{15}$ cm sec$^{-1}$.

\indent 4. {\it Accretion Rate}: Eddington rate is ${\dot M}_{Ed} = 1.44 \times 10^{17} 
\frac{M_{BH}}{M_\odot} $gm sec$^{-1}$. This rate is determined by equating inward 
gravitational force on protons and the outward radiative force on electrons (e.g. [1]).
${\dot M}_{Ed}$ is an upper limit of 
accretion rate, and is strictly valid for spherical flow on Newtonian
stars and when the Thomson scattering is dominant. Critical rate is  
${\dot M}_{Crit} = {\dot M}_{Ed} / \eta$, where, $\eta$ denotes the 
efficiency of energy extraction from infalling matter. $\eta \sim 0.06$ 
for Schwarzschild black holes and $\eta \sim 0.4$ for extreme Kerr black holes. 

\indent 5. {\it Luminosity}: Eddington luminosity is $L_{Ed}={\dot M}_{Ed} c^2=
1.3 \times 10^{38} \frac{M_{BH}}{M_\odot}$ ergs sec$^{-1}$. The critical luminosity 
is $L_{Ed}/\eta$. 

\indent 6. {\it Density of gas}: $\rho_g = \frac {{\dot M}_{Ed} t_g}{r_g}^3 
= 5.3 \times 10^{-5} (\frac{M_{BH}}{M_\odot})^{-2}$ gm $cm^3$.

\indent 7. {\it Virial Temperature}: $T_{virial}= \frac{1}{k} \frac {GM_{BH} m_p}{r_g} 
=5.2 \times 10^{12} $K.

\indent 8. {\it Black Body Temperature}: $T_{BB}  = (\frac{L_{Ed}}{\sigma r_g^2})^{1/4}
= 7.1 \times 10^7 (\frac{M_{BH}}{M_\odot})^{-1/4}$ K.

\ indent 9. {\it Magnetic fields}: Field strength is estimated from the assumption
of equipartition: $B_E = (2\pi c^4 m_p /\sigma_T G M_{BH})^{1/2} \approx 3 \times 
10^8 \frac{M_{BH}}{M_\odot}$ G.

\subsection{Effective Potential of Photons and Their Trajectories Around a 
Black Hole}

Photons orbit in null geodesics governed by the metric of the spacetime.
In Schwarzschild spacetime, the effective potential of photons is 
obtained from the null geodesic equation ($GM_{BH}/c^2$ is the length
unit):
$$
(\frac{dr}{d\lambda}) ^2  =E^2- \frac{l^2}{r^2} (1 - \frac{2}{r}) 
\eqno{(1a)}
$$
This can be rewritten as 
$$
\frac{1}{b^2}- \frac{1}{r^2} (1 - \frac{2}{r}) = \frac{1}{b^2}-
V_{phot}=(\frac{dr}{d\lambda^\prime}) ^2 
\eqno{(1b)}
$$
The nature of the potential $V_{phot}$ is shown in Fig. 1a. This shows that the
photons with an impact parameter $b=l/E < 3\sqrt{3}$ would be swallowed by
the hole. Here, $l={u_\phi}$ is the specific angular momentum and 
$E=-u_t$ is the specific energy of a photon. If $b\gsim 3\sqrt{3}$, 
photon would escape. It would be interesting to 
know about the trajectories of photon around a black hole in order 
to understand why it is difficult to detect it. The equation 
obeyed by photons around a Schwarzschild black hole is the null geodesic equation:
$$
(\frac{du}{d\phi})^2 = 2 u^3 - u^2 + \frac{1}{D^2}
\eqno{(2)}
$$
where, $u=1/r$, $D=L/E$, $l= u_\phi$.

Fig. 1b shows a collection of photon trajectories at different distances 
from a black hole. Shaded cone (`absorption cone') drawn at 
each radius indicates the directions in which photons are swallowed 
by the black hole. Photons emitted from the rest of the region (`emission 
cone') can escape to a large distance. Half angle $\psi$ of the
cones are given by $\psi=sin^{-1}\ 3\sqrt{3}/[r\sqrt(1-2/r)]$. 
For instance, only half of the photons emitted isotropically from 
a point source at $r=1.5r_g = 3GM_{BH}/c^2$ would escape to a 
large distance. This has a significant implications in spectral 
properties of black holes as will be discussed in Section 4. All the 
photons emitted from $r=r_g$ (dotted circle) are absorbed by the black hole.
Hawking radiation emitted from immediate vicinity of a black hole horizon
can come out due to quantum effects, but they will not be discussed here.
 
\subsection{Effective Potential of Particles Around a Black Hole}

Another interesting property of black holes which is often useful to identify 
them is the nature of the effective potential of a particle with a test 
mass. The effective potential of the particle is:
$$
V_{eff}=[(1+\frac{l^2}{r^2}) (1-\frac{2GM_{BH}}{r})]^{1/2} .
\eqno{(3)}
$$
The potential starts developing a minimum when $l\gsim 2\sqrt{3}$.
For $l<2\sqrt{3} GM_{BH}/c$ the potential poses no obstruction to incoming 
matter, and matter can fall on to a black hole as easily as a spherical
flow devoid of angular momentum. This is in sharp contrast with 
what is seen in a Newtonian geometry. Here, the effective potential is:
$$
V_{eff, N}=1-\frac{GM_{BH}}{r} + \frac{l^2}{2r^2} .
\eqno{(4)}
$$
which presents an infinitely high barrier to the flow with even an insignificant
angular momentum. Fig. 2 shows the potential barrier both around a 
Schwarzschild black hole
(solid curve) as well as around a Newtonian star (dashed) as a function
of the radial distance (measured in units of $GM_{BH}/c^2$). The solid
curves are drawn for (from bottom curve to the top curve) $l=0,\ 2,\ 3,\
l_{ms}=2 \sqrt{3}, \ l_{mb}=4$ \& $5$ respectively and the dashed curve is drawn for
$l=2$ (angular momenta are measured in units of $GM_{BH}/c$).
$l_{ms}$ is called the marginally stable angular momentum, since the
closed orbits are impossible below this $l$. At $r=r_{ms}= 3 r_g= 6 GM_{BH}c^2$, 
$V_{eff} (l=l_{ms})$ has a point of inflection, and is called the marginally 
stable orbit. This closed orbit is the last stable orbit nearest to the
black hole. For $r<r_{ms}$ matter cannot stay in a stable orbit and must dive
into a black hole even with a slighted perturbation. This is an important 
ingredient in constructing accretion disk models as the inner sonic point
where the flow becomes supersonic is close to this radius.

For $l_{mb}=4GM_{BH}/c$, $V_{eff}=1$ at $r_{mb}=2 r_g=4GM_{BH}/c^2$. In this
case, the orbit is marginally bound and the closest such orbit is located 
at $r_{mb}$. For $l \gsim l_{mb}$, $V_{eff} >0$ in some region, and the matter
will be bounced back to large distance. There is no bound orbit in region 
$r<r_{mb}$.

\subsection{Fundamental Properties of Black Hole Accretion}

Study of modern accretion processes on stars and compact objects began with 
the revolutionary work on spherical flows onto ordinary stars by Bondi [2],
although the `black hole' phrase was not known (at least in the context of
Astrophysical object) at the time of that publication. 
Bondi solution was obtained in Newtonian geometry for pointlike mass. The 
general conclusion was that the subsonic (i.e., radial Mach number $M= v/a_s < 1$,
where $v$ is the radial velocity and $a_s$ is the adiabatic sound speed of the
infalling matter)
flow with specific energy ${\cal E}\sim n a_\infty^2 \geq 0$ (where $n$  
is the polytropic index of the flow, and $a_\infty$ is the adiabatic sound 
speed at a large distance) which begins at rest at infinity would pass 
through a sonic point ($M=1$) and remains supersonic $M>1$
till the star surface where it stops and becomes subsonic. 
In this way, the boundary layer could be studied 
as a part of the inflow itself [3].
Alternatively, matter may remain subsonic throughout (e.g., [4]). 
For a black hole accretion, the flow passes through the horizon 
with the velocity of light, and therefore it must be supersonic
on the horizon. These conclusions are valid for rotating flows 
as well as for flows around rotating black holes [5].
For a recent review on spherical flows see, [4-6], and references therein.

With the Bondi solution in hand, the excessive luminosities of quasars 
and active galaxies in the sixties and seventies were readily 
conjectured to be due to gravitational energy release of matter 
accreting on black holes. However, it became apparent very soon that
rapidly inflowing spherical matter is of very low density and advects 
virtually all the energy (save a small loss due to bremsstrahlung) through 
the black hole horizon. Magnetic dissipation could increase the 
efficiency of emission [7-8], but the assumptions which 
went in (for instance, equipartition of gas and magnetic fields) 
were not at all satisfactory. This is because the magnetic field
and the gas are compressed at different rates: the pressure due 
to a radial field is compressed at $P_{mag} \propto r^{-4}$, 
while that due to adiabatic gas is compressed at $P_{gas} \propto 
r^{-5/2}$ (e.g., [9]). Equipartition achieved at 
any given radius is quickly off-balanced in another as the matter
moves in. A serious problem is that locally excess energy of the magnetic
field need not be dissipated within the disk itself, since the fields
cannot be easily anchored in a stable disk as in a star [10].
Shakura \& Sunyaev [11]
and Novikov \& Thorne [12] 
increased the efficiency of emission 
by assuming the flow to be rotating in Keplerian orbits just as 
in Saturn's ring. This basically rotating matter is of high density 
and the radiation emitted from this optically thick flow is roughly 
of black body type. The `multicolour' black body emission (obtained 
by summing black body contributions from a large number of annuli 
on the disk) roughly agrees with the observed accretion disk spectra 
in binary systems as well as in active galaxies (see, e.g., [6] for 
references). One novel, though very simplifying, assumption made in 
this disk model is that one could quantify the unknown viscosity
by using a so-called `$\alpha$' parameter. The viscous stress $W_{r\phi}$
is simply written as $-\alpha P$, where $P$ is the total pressure.
This disk model (apart from a few corrections here and there)
still remains the so-called `standard model' after twenty
five years of its introduction to the community. 

This Keplerian description of purely rotating disk, is probably a bit too
simplistic. Eardley \& Lightman [13] solved the two temperature
problem and pointed out that the Keplerian disk is viscously and 
thermally unstable if the viscosity parameter $\alpha$ 
is a constant throughout the disk. Even two decades ago,
observed spectra of the black hole candidates, such as Cyg X-1 [14]
indicated that the spectrum 
consists of two distinct components. The soft X-ray bump 
in these spectra could be explained by multicolour black 
body emission from a Keplerian disk. The power-law component 
of the spectra was explained by Comptonization of softer photons by
hot electrons from `Compton clouds', or, magnetic corona [15-19].
The softer photons originate 
from a Keplerian disk, and the origin of the Compton cloud 
including magnetic corona remained unspecified, until recently,
when it was realized that the so-called Compton cloud is probably the
inner part of the disk itself when the disk is described by 
advective flow [20-21].
The behaviour of the 
power-law component was complex: the energy spectral 
index $\alpha$ ($F_\nu \propto \nu^{-\alpha}$) apparently 
stays closer to $0.5-0.7$ when the soft bump is very weak 
or non-existent (and remains almost constant even when the 
intensity of the soft bump changes by a factor of several), 
and closer to $1.5$ when the soft bump is very strong. 
(Note that we are using the same customary notation $\alpha$ 
to denote the viscosity parameter as well as the energy 
spectral index. But this will not cause confusion.)
In the first case, most of the power is emitted in the 
hard component and the black hole is said to be in the 
`hard state'. In the second case, most of the power is 
emitted in the soft component and the black hole is said 
to be in `soft state' [20, 22-23 and references therein].
The energies of the so-called 
`soft' and `hard' radiations depend on the mass of the 
black hole. For a stellar black hole (typically resulting 
from a supernova explosion) of mass $10M_\odot$, the soft 
radiation bump would be in $\sim 1-3$keV, while for a 
super-massive black hole (typically resulting from  
protogalactic collapse at the center of galaxies) of 
mass $10^{8-9}M_\odot$, the soft radiation bump would 
be in the extreme ultraviolet (EUV) to ultraviolet (UV)
region. Hard radiations of $100$keV or even up to an MeV 
are not uncommon in both types of black holes. Typical 
Compton spectra using Sunyaev-Titarchuk [15]) type 
analysis suggests that in the low state the electron temperature 
of the hot region is roughly $100-200$keV for stellar 
mass black holes and around $40-60$keV for super-massive black holes. 

The nature of the emitted radiation is clearly very complex
and no simple solution such as the Bondi flow or a Keplerian
disk could explain it completely. In reality, of course, it
is difficult to form either of these flows anyway. Since the incoming flow must 
have some angular momentum as it is coming from an orbiting companion
(as in a binary system) or some orbiting stars (as in a galactic nucleus),
the flow cannot be purely Bondi-like. Similarly, the inner 
boundary condition on the horizon (that the flow velocity be equal 
to the velocity of light) suggests that the flow must be supersonic,
and hence sub-Keplerian at least close to the horizon [6, 24].
Again, unlike in a Keplerian disk, where Keplerian distribution 
is guaranteed whatever viscosity 
is present, in a realistic flow (which includes advection by default), the 
angular momentum distribution is to be determined self-consistently
from the transport equations. It may join a Keplerian disk 
at several tens to hundreds of Schwarzschild radii away depending on the specific 
energy, angular momentum and viscosity. Thus a realistic flow {\it must} 
be an intermediate solution between the Bondi flow and a Keplerian
disk. If the specific energy is positive, the advective flow would
have the ability to pass through two saddle type sonic points 
through shocks (but need not have shocks always, especially when the 
Rankine-Hugoniot conditions are not satisfied), but if the specific energy
is negative, it may pass through only one sonic point and join with
a cold Keplerian disk farther away. These conclusions are valid 
for {\it any} accretion rate or viscosity.  For higher accretion rates,
the advective flow should cool down due to  Comptonization, but
then the energy would come down to negative. Sustained magnetic heating,
could keep the energy to be positive and therefore flow can remain
advective. Far away from the black hole, the flow may be cold and Keplerian
(may even be sub-Keplerian if matter is accreted from a large number of
stars), but close to the black hole, the flow must be sub-Keplerian. Apart from 
this deviation of the angular momentum distribution, the geometrical shape
and internal dynamics of matter are also very much changed from a standard 
model. The presence of some angular momentum causes the centrifugal barrier 
supported geometrically thick region to be formed around a black hole whose
property is similar to that of a boundary layer. This we call CENBOL
(CENtrifugally supported BOundary Layer). The evolution
of the accretion disk model is schematically shown in Fig. 3(a-c): In 3(a)
the Bondi flow; in 3(b), the Keplerian disk and in 3(c), the generalized 
advective disk are shown. The figure in 3(c) is based on combination
of advective disk solutions in different regime. Beside each model, 
a typical spectrum is also schematically shown 
in $\nu log(F_\nu)$ vs. $log(\nu)$ scale. In (a), 
the spectrum is hard, mostly due to bremsstrahlung; in (b), the
spectrum is soft and multicolour blackbody type, and in (c), 
it is the combination of the soft X-ray and hard X-ray 
emitted from the disk. In 3(c), variation of the spectra
relative accretion rates in the Keplerian and sub-Keplerian components
(or, equivalently, viscosity variation in the disk) is shown where 
the transition from the hard to soft state is achieved. 

In this context, it is to be noted that advective accretion 
flows are those which self-consistently include advection 
velocity as in Bondi flows at the same time include rotation, viscosity, 
heating and cooling processes. For a black hole accretion, these 
are the same as viscous transonic flows (VTF) discussed in detail in 
Chakrabarti [5, 25].
Special cases of VTFs are the so called
slim disks for optically thick flow [26]
and Advection Dominated Flows for optically thin flow [27].
However, global solutions of these special cases were not found to be
satisfactory [5]. Thus, we use the more general advective disk solution.
For a neutron star accretion, the flow need not be transonic 
(e.g., could be subsonic everywhere) and the advective disks 
include that possibility as well. 

There are several models in the literature which were brought
in to explain different observational features from time to time.
These are far from self-consistent and serve only special purpose for 
which they were invoked. Foundation of the advective disk solution 
discussed here, on the contrary, is on the most general set of equations
from which all other models emerge as special cases. It is therefore
no surprise that this solution seems to have the `right features in the
right regime'. The Bondi flow and the Keplerian disks are both extreme 
cases: one does not have rotation and the other does not have advection 
properly included. The first one is basically
energy conserving (roughly valid for low accretion rate)
spherical advective flow and the second one is highly dissipative
(roughly valid for viscous, high accretion rate) disklike 
flow. Attempts to fill the gap, i.e., to find intermediate structures
were made since early eighties. Paczy\'nski and his collaborators
advanced models in two different directions: (1) Thick accretion disks 
[28] and references therein): where the flow 
is still rotation dominated, but non-Keplerian. The radiation 
pressure is high enough to hold matter vertically, preventing it from 
collapsing. This thick disk model was clearly valid for high accretion 
rate and the radiation pressure in the funnel was assumed to push
matter vertically to form outflows and jets. Rees et al. [29]
pointed out that the low accretion rate flow can support vertical structure
due to strong ion pressure while most of the energy is advected. 
These models were not globally complete as the disk is generally 
non-accreting. (2) Transonic flows [30-31]:
Here the radial motion is also included but global solutions were not
explored. Preliminary non-dissipative solutions indicated [32]
that unlike  Bondi flow, there are two saddle
type sonic points in an advective flow. Abramowicz \& Zurek [33]
further concluded that the same matter could go through the
outer (Bondi) or inner (disklike) sonic points, although it is now
known that they have different entropies and should be counted as
by two different flows unless connected by standing shocks (see, [5] for details.).
Matsumoto et al.  [34]
tried to mend the inner edge of the 
Keplerian disk (although the sonic point was chosen to be nodal) while
Abramowicz et al. [26], in the so-called slim disk tried to
find global solutions of the transonic flow in the high accretion
rate limit without success (although local solutions were seen to be
stable due to advection). The transonic solution of Fukue [35] had 
a standing shock wave very similar to those in winds by Ferrari et al. [36]. 
Chakrabarti [3, 5, 25]
found all possible topologies of advective disk
solutions in viscous and non-viscous flows, with and without magnetic
fields. The recently re-discovered ion tori [29]
(dubbed as advection dominated flow) by  [27, 37]
is as ad hoc as its predecessor since it is assumed to behave
something like `corona without a disk'. This corona comes about 
by inexplicable evaporation method of the underlying Keplerian 
disk at a low accretion rate. The spherical distribution of matter that comes
about (see Fig. 1 of Narayan [38]) even when angular momentum is
included is equally inexplicable. (This controvertial
solution or its consequences are clearly irrelevant in black hole astrophysics
and will not be discussed in  this review any farther.)
The post-shock and/or the CENBOL regions of advective disk solutions
[3, 5, 25]
resemble those of thick disks (but accreting!). High accretion rate solutions 
of advective disk are the globally correct solutions of `slim disks' (unlike
Abramowicz et al., [26] model, these advective flows have angular momentum
correctly approaching a Keplerian disk) and low accretion
rate advective disks are the correct `advection dominated flows'
(unlike  Narayan \& Yi, [27, 37, 38]
model, these flows naturally deviate
from a Keplerian disk specially for low viscosity and positive specific 
energy and NOT for high viscosity, and no evaporation is required). 
Recently, some groups are attempting to try to reproduce advective 
disk solutions in their respective regimes [39-40].
The success has been limited because
of less general approach of solving the eigenvalue problem that
the equations posit. In their approach the flow is `let loose'
from a Keplerian disk at an arbitrary distance 
with an arbitrary initial radial and angular velocity components
and only one sonic point is assumed (thus by construction
this method cannot find shocks, for instance, even when the 
specific energy is positive). In the advective disk approach [3, 5, 25]
the flow is also allowed to have `discontinuous' 
(shock) solutions which joins two branches passing through two 
sonic points. When such joining is not possible, the inflow shows 
non-stationary behaviour, unless only one sonic point is present to 
begin with.  With viscosity variation, the net accretion flow
is Keplerian disklike in some region, and advective in some other
region. Today, the need for having the admixture of Keplerian and 
sub-Keplerian components (as first quantitatively pointed out by [20])
in the accretion flows is clearly recognized in most of the observations.

An accretion disk must fall in and advect. It was on this philosophy
Bondi flow was studied originally. In the intervening period, specially,
in the seventies and early eighties, rotating Keplerian disk took over
while inclusion of advection was considered to be a `new model'! 
The question of whether an advective disk can explain observations
is outright irrelevant because this disk represents the only
self-consistent solution of the governing equations which are derived
from fundamental laws of nature (such as conservation of energy and momentum).
The same solution produces boundary layer of black holes 
(specifically for hot flows) and neutron stars. The importance of 
CENBOL was not appreciated originally, but now they are indispensable 
in most explanations, given that they obviate the need to 
construct ad hoc `Compton Clouds' (e.g., [41-42],
and references therein). The future of the black hole 
astrophysics is most certainly the correct understanding 
of the CENBOL region of the flow.

In order to establish the general behaviour of 
matter described above, we now present all the possible
solutions of non-self-gravitating test flow around a Kerr
black hole. We use $t$, $r$, $\theta$ and $z$ as the coordinates. 
We choose the geometric units where $G=M_{BH}=c=1$ ($G$ is the 
gravitational constant, $M_{BH}$ is the mass of a black hole and 
$c$ is the velocity of light). We also consider $|\theta-\pi/2| <<1 $
for a thin flow on the equatorial plane in vertical equilibrium.
We consider a perfect fluid with the stress-energy tensor,
$$
T_{\mu\nu} = \rho u_\mu u_\nu + P(g_{\mu\nu} + u_\mu u_\nu)
\eqno{(5)}
$$
where, $P$ is the pressure and $\rho=\rho_0(1+\pi)$ is the 
mass density, $\pi$ being the internal energy. We ignore the self-gravity 
of the flow as well as the contribution due to viscous
dissipation. We assume the vacuum metric around a Kerr black hole to be of 
the form [12],
$$
ds^2 = g_{\mu\nu}dx^\mu dx^\nu= -\frac{r^2 \Delta}{A} dt^2 + \frac{A}{r^2}
(d\phi-\omega dt)^2 +\frac{r^2}{\Delta} dr^2 + dz^2
\eqno{(6)}
$$
Where,
$$
A= r^4 + r^2 a^2 + 2 r a^2
\eqno{(7a)}
$$
$$
\Delta= r^2 - 2 r + a^2
\eqno{(7b)}
$$
$$
\omega = \frac{2 a r}{A} .
\eqno{(7c)}
$$
Here,  $g_{\mu\nu}$ is the metric coefficient and $u_\mu$ is the 
four velocity components:
$$
u_t= - \left[\frac{\Delta}{(1-V^2)(1-\Omega l)(g_{\phi\phi}+l g_{t\phi})
}\right]^{1/2}
\eqno{(8a)}
$$
and
$$
u_\phi=-l u_t
\eqno{(8b)}
$$
where, the angular velocity is
$$
\Omega=\frac{u^\phi}{u^t}=-\frac{g_{t\phi}+ l g_{tt}}{g_{\phi\phi}+l g_{t\phi}}
\eqno{(9)}
$$
and $l=- u_\phi/u_t$ is the specific angular momentum. The radial velocity $V$ in 
the rotating frame is (see, [25] p. 137)
$$
V=\frac{v}{(1-\Omega l)^{1/2}}
\eqno{(10)}
$$
where,
$$
v=(-\frac{u_ru^r}{u_tu^t})^{1/2}.
\eqno{(11)}
$$
It is trivial to check that $V$ above is unity on the horizon independent
of the initial condition.

Since even for the extreme equation of state $P=\frac{1}{3} \rho_0 c^2$,
($p$ is the isotropic matter pressure and $\rho$ is the rest mass
density), the speed of sound is $a_s = \frac{1}{\sqrt {3}}$, the 
Mach number of the flow is $M = v/a_s = \sqrt{3} >1$, any physical
flow must be {\it supersonic} on the horizon. This inner 
boundary condition has a profound effect on the structure of the 
accretion disk around a black hole as we will see here.

In the present review, we shall concentrate on the time independent 
solutions of the underlying hydrodynamic equations.
The equation for the balance of the radial momentum is obtained from
$T^{\mu\alpha}_{;\alpha}=0$:
$$
\vartheta\frac{d\vartheta}{dr} + \frac{1}{r \Delta} [a^2- r + 
\frac{A\Gamma^2 B}{r^3}]\vartheta^2 + \frac{A \Gamma^2}{r^6} B 
+ (\frac{\Delta}{r^2} + \vartheta^2 )\frac{1}{p+\rho} \frac{dP}{dr}=0
\eqno{(12)}
$$
where,
$$
\Gamma^2= [1-\frac{A^2}{\Delta r^4} (\Omega - \omega)]^{-1} ,
\eqno{(13)}
$$
$$
B=(\Omega a - 1)^2 - \Omega^2 r^3 ,
\eqno{(14)}
$$
and
$$
\vartheta=u^r.
\eqno{(15)}
$$
Here and hereafter a comma is used to denote an ordinary derivative and a 
semi-colon is used to denote a covariant derivative.
The baryon number conservation equation (continuity equation)
is obtained from $(\rho_0 u^\mu)_{; \mu}=0$ which is,
$$
\noindent {\dot M}= 2 \pi r \vartheta \Sigma= 2\pi r \vartheta \rho_0 H_0
\eqno{(16)}
$$
where,
$$
H_0= (\frac{P}{\rho_0})^{1/2} \frac{r^{3/2}}{\Gamma} 
[\frac{(r^2+a^2)^2- \Delta a^2}{(r^2 + a^2)^2 + 2 \Delta a^2}]^{1/2}
\eqno{(17)}
$$
is the height of the disk in vertical equilibrium [12].
The equation of the conservation of angular momentum 
is obtained from $(\delta^{\mu_\phi} T^\prime_{\mu \alpha})_{;\alpha}=0$, 
and one obtains,
$$
\rho_0 u^\mu (h u_\phi)_{,\mu}= (\eta \sigma_\phi^\gamma)_{;\gamma}
\eqno{(18)}
$$
where,
$$
\eta=\nu\rho_0
\eqno{(19)}
$$
is the coefficient of dynamical viscosity and $\nu$ is the coefficient of
kinematic viscosity. The stress-energy tensor $T^\prime_{\mu\nu}$
contains the term due to viscous dissipation (e.g., [43]).
When the rotation is dominant ($\vartheta <u^\phi$),
the relevant shear tensor component $\sigma_\phi^r$ is given by [44],
$$
\sigma_\phi^r= - \frac{A^{3/2} \Gamma^3 \Omega_{,r} \Delta^{1/2}}{2r^5}
\eqno{(20)}
$$
so that the angular momentum equation takes the form [24],
$$
{\cal L}-{\cal L}_{+}=  -\frac{1}{\vartheta r^5}\frac{d\Omega}{dr} \nu A^{3/2} 
\Gamma^3 \Delta^{1/2}.
\eqno{(21)}
$$
Here, we have corrected the  angular momentum transport
equation of Novikov \& Thorne [12] as derived in [24]. In this equation on the
left hand side, the fluid angular momentum ${\cal L}=h u_\phi$ [24] rather 
than particle angular momentum $u_\phi$ [12] has been used. For an inviscid 
flow, $\eta=0$ one recovers ${\cal L}=constant$ as it should be. Similarly, the 
radial velocity term is included (eq. 12) and angular momentum is allowed to be 
non-Keplerian (eq. 21). ${\cal L}_+$ is the angular momentum on the horizon 
since the {\it rotational} shear (as defined by eq. 20) vanishes there.
In presence of significant radial velocity, the shear in eq. (20)
is to be replaced by its full expression, $\sigma^{\mu\nu}=(u^\mu_{;\beta} 
P^{\beta\nu} + u^{\nu}_{;\beta}P^{\beta\mu})/2 - \Theta P^{\mu\nu}/3$ 
where $P^{\mu\nu}= g^{\mu\nu}+u^\mu u^\nu$ is the projection tensor 
and $\Theta= u^\mu_{;\mu}$ is the expansion  [1].

Entropy generation equation is obtained from the first law of thermodynamics
along with the baryon conservation equation $(S^\mu)_{;\mu}=
[2\eta \sigma_{\mu\nu}\sigma^{\mu\nu}]/T - Q^-$ :
$$
\vartheta \Sigma (\frac{dh}{dr} - \frac{1}{\rho_0}\frac{dp}{dr}) =  Q^+-Q^- =
2\nu \Sigma \sigma_{\mu\nu}\sigma^{\mu\nu} - Q^ -
\eqno{(22)}
$$
where $Q^+$ and $Q^-$ are the heat generation rate (by viscosity, 
exothermic nuclear energy generation, magnetic dissipation, etc.) and the heat loss rate (by 
radiative cooling, by endothermic reactions, etc.) respectively. $h$ is 
the specific enthalpy: $h=(p+\rho)/\rho_0$. Here, the terms contributed by 
radiation have been ignored as well. Using rotational shear 
as given in eq. (20), the entropy equation takes the form,
$$
\vartheta \Sigma (\frac{dh}{dr} - \frac{1}{\rho_0}\frac{dp}{dr}) = 
\frac{\nu \Sigma A^2 \Gamma^4 (\Omega_{,r})^2}{r^6} - Q^- .
\eqno{(23)}
$$
Of course, for accuracy, one should use the full expression for $\sigma_\phi^r$.

This set of equations are solved simultaneously keeping in mind that the 
shock waves may form in the flow, especially when the specific energy is
positive. The following momentum balance condition [24],
$$
W_-n^\nu + (W_-+\Sigma_{0-}) (u_-^\mu n_\mu)u_-^\nu  =
W_+n^\nu + (W_++\Sigma_{0+}) (u_+^\mu n_\mu) u_+^\nu
\eqno{(24)}
$$
along with the conservation of energy and mass fluxes (together, these 
conditions are known as the Rankine-Hugoniot conditions) must be fulfilled
at the stationary shock. Here, $n_\mu$ is the four normal vector 
component across the shock, and $W$ and $\Sigma$ are vertically 
integrated pressure and density on the shock surface. Here, the subscripts 
$-$ and $+$ denote the pre-shock and post-shock quantities respectively. 

The equations presented above are applicable to optically thin as well as
optically thick flows for any general heating and cooling processes. 
For  a given viscosity prescription and the exact 
cooling processes (depending on the optical depth of the flow), it is 
usual to reduce the above set of equations in the form:
$$
\frac{du}{dr}=\frac{N}{D}
\eqno{(25)}
$$
where, $N$ and $D$ are the smooth functions of radial coordinate (unless
there are non-linearities which prevent such a reduction. In that case
sonic curve analysis is done, see, Castor, Abott \& Klein [45]
in Newtonian context and Flammang, [46] in general relativistic
winds). The procedure of obtaining the complete solution 
is then similar to what is presented in obtaining the global 
solutions of viscous transonic flow using pseudo-Newtonian 
potential ([5, 25] and references therein). We shall present 
the complete set of solutions of these equations shortly. Before
doing so, we prove two fundamental points about the advective flow.

\subsubsection{Flow Must be Sub-Keplerian on the Horizon}

We first  rewrite $B$ of equation (12) as
$$
B= -(\Omega - \Omega_{K+}) (\Omega - \Omega_{K-})
\eqno{(26)}
$$
where, 
$$
\Omega_{K\pm} = \pm \frac{1}{r^{3/2} \pm a}
\eqno{(27)}
$$
are the Keplerian angular momenta for the co- (plus sign) and 
contra- (minus sign) rotating flows. From eqn. (12) one notes 
that since $u^\prime <0$ and $p^\prime \sim 0$ on the horizon, 
one must have,
$$
B>0
\eqno{(28)}
$$
on the horizon, i.e., $\Omega <\Omega_{K+}$, or $l <l_{Kep}$. This can also
be shown  more generally context by computing $a_s^2$ at the sonic
point and imposing the condition that $a_s^2 >1$. It is seen that
at the sonic point the flow must be sub-Keplerian. 
In the inviscid flow, this means that the sonic point should
form only in sub-Keplerian region. In presence of viscosity, since angular
momentum transport rate is almost zero close to the horizon 
and the distribution is almost flat (see, Fig. 6), the flow
would maintain the sub-Keplerian  nature between the inner 
sonic point and the horizon. 

\subsubsection{Rotating Flow Must Have a Centrifugal Barrier}

An important ingredient of the state-of-the-art accretion flow is the
centrifugal pressure supported denser region close to a black hole.
Roughly speaking, the infall time scale being very short compared 
to the viscous (transport of angular momentum) time scale, the angular
momentum $l$ remains almost constant close to the black hole 
particularly for lower viscosity. As a result, the centrifugal 
force $l^2/r^3$ increases much faster compared to the gravity 
$\sim 1/r^2$ as the flow approaches the black hole. Matter starts 
piling up behind this centrifugal barrier and becomes denser, 
with opacity $\tau \sim {\dot m}$, where ${\dot m}$ is the 
accretion rate in units of the Eddington rate. Eventually, of 
course, the gravity wins and the matter enters into the black hole
supersonically. Since the effective potential turns over
for any angular momentum, matter with any amount of  angular 
momentum can be made to accrete on a black hole if it is
`pushed' hard enough. This is to be contrasted with the fact
that an infinite force is required to push matter to the surface of a 
Newtonian point mass with even an insignificant angular momentum
(Fig. 2). This is why a rotating flow has a saddle type sonic point
close to a black hole, while the closest sonic point for a Newtonian 
rotating flow is of unphysical `center' type. This will be demonstrated  below.

At the centrifugal pressure supported barrier (CENBOL), matter slows 
down and its thermal energy increases. In some region of the parameter space 
this slowing down takes place rather abruptly at a standing shock. 
Most of the thermal energy of the flow could be extracted from 
this region through inverse Compton effect if soft photons are injected 
here from the Keplerian disk component. Whereas the boundary layer of a 
white dwarf is of thickness less than a percentage of its radius, 
the thickness of the boundary layer (CENBOL) of a black hole
is several (typically, $10-20$) times larger than its radius! If the neutron 
star is not compact enough (i.e., not within the inner sonic point 
of the flow), its boundary layer would also be of similar size. For 
compact neutron stars, the boundary layer could be very thin
because the shock transition ($x_{s1}$, or $r_{s1}$ in the notation of [3]) 
just outside the hard surface is allowed, unless the 
entire flow is subsonic. Absence of a centrifugal barrier
in a Bondi flow causes the flow to be inefficient and one requires
sufficient magnetic field to enhance the cooling efficiency [7, 8].
It is not to say that such enhancement should not take place, or, is not completely 
required in advective disks in explanation of spectra 
(especially at high energies when synchrotron
radiation plays a major role), but so far, the inclusion of the 
magnetic fields has not been done consistently. In a self-similar flow,
one equipartition of gas and magnetic field is achieved, it is 
always maintained. But the flow is hardly self-similar, with sonic points,
possible shocks, Keplerian flow boundaries etc. The major problem lies
is deciding what fraction of the magnetic field could be used for 
heating electrons (to enter in the $Q^+$ term in eq. 22), given that
most of the excess field may be expelled away in absence of 
adequate anchoring of the field on the disk [10].

CENBOL happens to have just the right set of properties: the  efficiency 
of its emission is neither almost zero as in a Bondi flow, nor fixed and maximum 
as in a Keplerian disk. Its size and optical depth are determined 
by viscosity and accretion rates and therefore give rise to varieties 
of spectral properties as are observed. The present review primarily 
emphasizes ways to identify black holes and therefore, ways in which 
CENBOL manifests itself in various observational phenomena. We discuss 
extensively how the spectral properties, both steady and non-steady, 
soft state and hard state, may be dependent on the properties of the 
CENBOL. We also show that as spin-offs, it may help supplying matter to 
the cosmic radio jets, explain metalicity of the galaxies 
and a host of other effects. 

\subsection{All Possible Ways to Dive into a Black Hole}

We now present all possible ways matter can enter into a black hole.
In obtaining a global solution one supplies the conserved
quantities at the inner or the outer (e.g., Keplerian or 
sub-Keplerian flows injected at the outer region) boundary, 
depending upon whether one is interested in the wind 
solution or the accretion solution. For a given angular momentum 
$l$, the remaining unknowns are $V (r)$ and $a_s (r)$. But one requires
only one extra boundary condition, e.g., ${\cal E}$, since
two sonic point conditions (eqs. 31a and 31b) introduce only one extra
unknown, namely, $r_c$. Thus, the supply of the initial
specific energy ${\cal E}$ and the specific angular momentum $l$
are sufficient for a complete solution from the horizon to infinity.
For a viscous flow, one clearly has to supply the distribution of
viscosity $\eta(r)$ (e.g., ion or magnetic viscosity) itself. Simple
viscosity prescription [11] may not be very useful since that 
stress $-\alpha P$ is always negative, while in a general relativistic 
flow stress can change sign [24, 44].
the total pressure $P+\rho V^2$ (i.e., including ram pressure) is 
more appropriate than just $P$, especially when advection is 
significant [47].
Note that by definition, $\Omega=\omega$ on the horizon and thus the flow co-rotates 
with the black hole. Instead of specifying various quantities at the 
flow boundary, one can alternatively specify the location of a critical 
point along with the energy or the angular momentum [25].

In much of the parameter space, the flow is expected to be smooth
as in a Bondi flow. If the angular 
momentum is significant, matter can pile up behind the  centrifugal 
barrier close to the black hole and form a standing shock wave
where several quantities are actually discontinuous. At the shock, 
apart from the continuity of energy and mass flux,
the relativistic momentum balance condition (eq. 24) must be satisfied. 
Using Newtonian definition of the vertical integration (since thin flows
are being dealt with here) as in [3] and the definition of entropy 
accretion rate ${\dot {\cal M}}$, one easily finds that at the shock, the
following quantity:
$$
\Pi = \frac{\left [ \frac{a_{s}^2}{1-na_{s}^2} \right ]^{n+3/2}
\left ( \frac{2}{3\Gamma-1} + \frac {V^2}{a_s^2 (1-V^2)} \right )}
{{\dot{\cal M}}} 
\eqno{(29)}
$$
should be continuous. At the shock, entropy is generated 
(turbulent or other viscosities operating at 
the shock) which is then advected through the inner sonic point.

In this context, it is important to point out that eq. (29) 
is valid only for Rankine-Hugoniot shocks where the energy flux
is continuous. In an astrophysical flow which is open to surroundings, 
this need not be so and both energy and entropy could be lost to the 
surroundings. It has been estimated that one can release a burst of 
photons at the shock which could contain as high as a couple of 
percents of the rest mass energy [48, 49].
This is important. Potentially releasable energy even for an optically 
thin flow may not be released at all and may be completely advected 
towards the black hole.

The considerations mentioned above are valid for 
object whose external spacetime is similar to that of a Kerr black hole.
On a neutron star surface, however, matter has to stop and
corotate with the surface velocity. The inner boundary condition is therefore
sub-sonic. On a black hole, the flow must enter through the
horizon with the velocity of light, and therefore must be supersonic.
The supersonic flow becomes subsonic at the shock
and again becomes supersonic before entering through the horizon. 
Clearly, the flow has to become supersonic, before forming the shock 
as well, and therefore pass through another sonic point
at a larger distance away from the black hole. Thus, as a whole,
the flow may deviate from a hot Keplerian disk and (a) enter through the
inner sonic point only, or, (b) enter through the outer sonic point only, or, 
(c) pass through the outer sonic point, then a shock, and finally through an
inner sonic point if the shock conditions are satisfied. 
If the angular momentum is too small, then the flow has only one sonic point
and shocks cannot be formed as in a Bondi flow. 

In Fig. 4, the {\it entire} parameter space is classified according to 
the type of inviscid solutions that is prevalent [24].
The Kerr parameter $a=0.5$. (For classification of flows in pseudo-Newtonian 
geometry, see, [3, 21, 25]).
The adiabatic index $\gamma=4/3$ 
has been chosen. In the central box, the parameter space spanned by ($l, {\cal E}$) is divided 
into nine regions marked by $N$, $O$, $NSA$, $SA$, $SW$, $NSW$, $I$,
$O^*$, $I^*$. The horizontal line at ${\cal E}=1$ corresponds to the rest
mass of the flow. Surrounding this parameter space, we plot various
solutions (Mach number $M=v_r/a_s$ vs. logarithmic radial distance
where $v_r$ is the radial velocity and $a_s$ is the sound speed) marked
with the same notations (except $N$). Each of these solution topologies has
been drawn using flow parameters from the respective region of the central
box. The accretion solutions have inward pointing arrows and the 
wind solutions have outward pointing arrows. The crossing points are `X' 
type or saddle type sonic points and the contours of circular topology are 
around `O' type sonic points. If there are two `X' type sonic points, the 
inner one is called the inner sonic point and the outer one is called the 
outer sonic point. The solutions from the region `O' has only the outer sonic point.
The solutions from the regions $NSA$ and $SA$ have two `X' type sonic points
with the entropy density $S_o$ at the outer sonic point {\it less} than the
entropy density $S_i$ at the inner sonic point. However, flows from $SA$
pass through a standing non-dissipative shock since the Rankine-Hugoniot
condition is satisfied. The entropy generated at the shock $S_i-S_o$ 
is advected towards the black hole to enable the flow to pass
through the inner sonic point. Rankine-Hugoniot condition is not satisfied
for flows from the region $NSA$. Numerical simulations indicate [50]
that the flow from this region is 
very unstable and exhibit periodic changes in emission properties as the flow
constantly tries to form the shock wave, but fails to do so. Thus,
it is possible that the solutions are inherently time-dependent
(just as a simple harmonic oscillator)
in this region. The solutions from the region $SW$ and $NSW$ are very similar 
to those from $SA$ and $NSA$. However, $S_o \geq S_i$ in these cases.
Shocks can form only in winds from the region $SW$. The shock condition is not
satisfied in winds from the region $NSW$. This may make the $NSW$ winds
unstable as well, but the accretion through the inner sonic point is stable. A flow from 
the region $I$ has the inner sonic point and thus can form shocks (which 
require the presence of two saddle type sonic points) only if the inflow
is already supersonic due to some other physical processes. Each solution 
from regions $I^*$ and $O^*$ has two sonic points (one `X' and one `O')
and neither of them produces a complete and global solution. The region $I^*$
has an inner sonic point but the solution does not extend subsonically
to a large distance. The region $O^*$ has an outer sonic point, but the
solution does not extend supersonically to the horizon! When a significant 
viscosity is added, the closed topology of $I^*$ opens up and then the flow 
joins with a cool Keplerian disk [5, 25] which has ${\cal E} <1$. 
These special solutions of viscous transonic flows should not have shock 
waves. However, hot flows deviating from a Keplerian disk or sub-Keplerian 
companion winds can have ${\cal E}>1$ or, cool flows can be subsequently 
energized by magnetic flares (for instance). These could have standing 
shock waves as discussed above. Energetically, the flow should have 
${\cal E} >0$ as in a Bondi flow, to pass through the outer sonic
point, which is a pre-requisite to form standing shocks.

In [3] and [25], it was found that shock conditions
were satisfied at four locations: $r_{s1},\ r_{s2},\ r_{s3}$ and $r_{s4}$,
though $r_{s1}$ and $r_{s4}$ were found to be
not useful for accretion on black holes. Out of $r_{s2}$ and $r_{s3}$,
it was shown that $r_{s3}$ is stable for accretion flow  and $r_{s2}$
is stable for winds ([51] also see, [49, 52-54].
So we have plotted only $r_{s3}$ here in $SA$ and $r_{s2}$ in $SW$
solutions. Here, $o$ and $i$ are the outer and inner sonic points
respectively. In the solution from $SA$ (upper left box in Fig. 4), 
we chose $a=0.5$, $l=3$, ${\cal E}=1.003$. For these parameters, the 
eigenvalue of the critical entropy accretion rates at the two saddle type 
sonic points are ${\dot{\cal M}}_i=2.74 \times 10^{-05}$ and
${\dot{\cal M}}_o=1.491 \times 10^{-05}$ respectively. Here, 
${\dot{\cal M}}_o < {\dot{\cal M}}_i $, hence the flow through the 
outer sonic point joins the horizon with infinity (single arrowed curve).
The flow forms a shock and jump onto the branch which passes through
$i$ as shown by double arrows. The stable shock (shown by a vertical 
dashed line) is located at $a_3=r_{s3}=32.29$ (in notation of [3]). 
Only this jump, namely, a generation of entropy of amount 
${\dot{\cal M}}_i- {\dot{\cal M}}_o$ is allowed in order that 
the transonicity of the post-shock flow is guaranteed. The entropy 
generated at the shock is advected through the inner sonic point.
The optically thin flow is inefficiently cooled, which keeps the energy of 
the flow constant. This makes the flow much hotter than a Keplerian disk.
(This is typical of advective disks. See, [3], [25], [55].)
In $SW$ solution, we chose $a=0.5$, $l=3$, ${\cal E}=1.007$. For these 
parameters, the eigenvalue of the critical entropy accretion rates at 
the two saddle type sonic points are ${\dot{\cal M}}_i=3.12 \times 
10^{-05}$ and ${\dot{\cal M}}_o=5.001 \times 10^{-05}$ respectively. Here, 
${\dot{\cal M}}_i < {\dot{\cal M}}_o $, hence the flow 
through the inner sonic point $i$ joins the horizon 
with infinity (single arrowed curve). 
The accretion flow branch can no longer form a shock. But a wind,
first passing through $i$ can, as shown in double arrows. The stable shock
(shown by a vertical dashed line) is located at $w_2=r_{s2}=6.89$ in this 
case. Only this jump, namely, a generation of entropy of amount
${\dot{\cal M}}_o- {\dot{\cal M}}_i$ is allowed at the shock
in order that it can escape to infinity through the outer
sonic point $O$. This consideration, along with the continuity of
$\Pi$ (Eq. 19) allows one to locate stationary shock waves in a flow.
Note that though the flow has a shock-free solution (passing through 
$o$ for accretion in $SA$ solution and through $i$ for winds in $SW$ 
solution in Fig. 4), the flow would choose to pass through a shock because 
the latter solution is of
higher entropy. This fact has been verified through numerical simulations
of accretion and wind flows  [51, 56].
It is to be noted that the angular momenta associated 
with solutions which include shocks are not arbitrarily large. Rather, they
are typically less than the marginally stable value $l_{ms}$ as indicated 
in Fig. 4.

Global solutions which contain shock waves are not isolated solutions, 
but are present for a large range of energy and angular momentum. 
In Fig. 5, the variation of shock locations as a function of specific 
energy ${\cal E}$ is shown. Each set of curves, drawn for various specific 
angular momentum (marked on the set), consists of four segments: two for 
accretion ($a_2=r_{s2}$ and $a_3=r_{s3}$) and two for winds ($w_2=r_{s2}$ 
and $w_3=r_{s3}$). As discussed above, $a_2$ and $w_3$ (dotted curves) 
are unstable while $a_3$ and $w_2$ (solid curves) are stable. Kerr 
parameter $a=0.5$ is chosen. This example shows that stable shocks can form 
for a very wide class of flows. For corotating flows, the marginally stable
and marginally bound angular momenta are $l_{ms}=2.9029$ and $l_{mb}=3.4142$ 
respectively. Thus the shocks form for angular momentum around these values. 
Since centrifugal barrier becomes stronger with angular momentum, shocks are 
located at larger radii for higher angular momenta. Another important point 
to note is that the shock location increases when the specific energy is 
increased. In a quasi-spherical flow, with the same input radial velocity 
and angular momentum, the potential energy decreases with height (since 
the gravity becomes weaker), thereby increasing the specific energy
and therefore shocks bend backward with height. General statements 
made for the shocks are valid for the CENBOL as well, where shock itself 
is not formed, but the density and velocity variation follow the same 
general pattern.

The solution branch which is supersonic close to the axis is valid for
black holes while the solution branches subsonic close to the axis
are valid for neutron stars. This is discussed in details in [3]
and more recently in [4]. The solution entering through the horizon 
is unique, since it must pass through the sonic point. This is physically 
appealing since the properties of the horizon are independent of any physical 
parameters such as temperature and pressure  of the gas etc. The solution 
touching a neutron star surface is not unique in the same token since any 
number of subsonic branches from infinity can come close to the axis (either 
through shocks or without shocks). Of course, ultimately, the one which matches
with the surface properties of the star will be selected. In a black hole 
accretion, such choices are simply not present.

\subsection{Solution topologies of Viscous Flow}

When viscosity is present, three (instead of two) flow parameters govern the topology of the 
flow: the $\alpha$ parameter [11] or, its modified value $\alpha_\Pi$ 
in presence of advection [47], which determines the viscosity, 
the location of the inner sonic point $x_{in}$ through which matter must pass 
through, and the specific angular momentum $\l_{in}$ of the matter at 
the horizon (or, alternatively, that at $r_{in}$). It so happens that 
these parameters are sufficient to completely determine the solution.
Unfortunately, majority of the works in black hole astrophysics
has been done, not by using full fledged general relativity, but 
using pseudo-Newtonian geometry. Paczy\'nski \& Wiita [28]
pointed out that outside the Schwarzschild black hole, the spacetime
may be described using Newtonian equations, but changing the $-GM_{BH}/r$
potential to $-GM_{BH}/(r-2GM_{BH}/c^2)$ potential. There are small
deviations when fully general relativistic and Pseudo-Newtonian calculations 
are compared, but for all practical purposes the deviations are tolerable. The
{\it nature} of classification of the parameter space as well as the {\it 
nature} of the variation of shock locations as described above are valid for
for all Kerr parameters and is independent of the flow model that is employed 
as long as $\gamma <1.5$ [5, 25] though details vary from model to model 
[note that in [21], `$\gamma<1.5$' was misprinted as `$\gamma>1.5$']. 
For $\gamma >1.5$, two sonic points are not present in an adiabatic flow 
(but may be present if external heating is included for instance) and 
therefore the shocks cannot form, but the centrifugal barrier would still
exist as described in [5].

Typical hydrodynamic equations which govern vertically averaged advective 
flows in the pseudo-Newtonian geometry are as follows [5],

\noindent (a) The radial momentum equation: 
$$
v\frac{dv}{dx} + \frac{1}{\rho} \frac{dp}{dx}  + \frac {\lambda^2_{Kep} - 
\lambda^2}{x^3} = 0
\eqno{(30a)}
$$
\noindent (b) The continuity equation:
$$
\frac{d}{dx}(\Sigma x v) = 0
\eqno{(30b)}
$$
\noindent (c) The azimuthal momentum equation:
$$
v\frac{d\lambda(x)}{dx} - \frac{1}{\Sigma x} \frac{d}{dx} (x^2 W_{x\phi}) = 0
\eqno{(30c)}
$$
\noindent (d) The entropy equation:
$$
\Sigma v T \frac{ds}{dx} = \frac{h(x) v}{\Gamma_3 - 1}(\frac{dp}{dx} -
\Gamma_1 \frac{p}{\rho}) = Q^+_{mag}+Q^+_{nuc}+Q^+_{vis}-Q^-=
Q^+ - g(x, {\dot m}) q^+ = f(\alpha, x, {\dot m}) q^+ .
\eqno{(30d)}
$$
Here, we have included the possibility of magnetic heating (due to stochastic field)
and nuclear energy release as well.
On the right hand side, we wrote $Q^+$ collectively proportional to the
cooling term for simplicity (purely on dimensional grounds). The quantity
$f$ is almost zero on the Keplerian disk and may be about $1$ close to the horizon
(unless Comptonization is included which drains energy out of this region). Here,
$$
\Gamma_3 = 1+\frac{\Gamma_1 - \beta}{4 - 3\beta} ; \Gamma_1 = 
\beta + \frac{(4 - 3\beta)^2 (\gamma -1)}{\beta  + 12 (\gamma - 1)(1-\beta)}
\eqno{(31)}
$$
and $\beta(x)$ is the ratio of gas pressure to total (gas plus magnetic 
plus radiation) pressure:
$$
\beta(x) = \frac {\rho k T/\mu m_p}{\rho k T/\mu m_p + {\bar a} T^4/3 +
B(x) ^2/4\pi}
\eqno{(32)}
$$
Here, the radial distance $x$ is in units of $2GM_{BH}/c^2$,
$B(x)$ is the strength of magnetic field in the flow, $p$ and $\rho$
are the gas pressure and density respectively, $\Sigma$ is the density
integrated in vertical direction, $T$ is the temperature of the flow 
(proton and electron), $h(x)$ is the height of the flow chosen to be 
in vertical equilibrium, ${\bar a}$ is the Stefan's constant, $k$ 
is the Boltzmann constant, $\mu$ is the electron number per particle 
(and is generally a function of $x$ in case of strong nucleosynthesis 
effects), $m_p$ is the mass of the proton. 
Two temperature solutions are important in the case where strong cooling
is present [20]. In an optically thick gas, the cooling is governed by black body
emission, while in optically thin limit it could be due to bremsstrahlung,
Compton effects, synchrotron radiation etc. (see, [57]).
Except for Compton scattering, other coolings are computed analytically
and is very simple to take care of. A novel method to include Compton 
cooling in accretion flows (first used in [20]) is to fit analytical 
curves of the numerical results of Sunyaev \& Titarchuk [16] for 
the cooling function as a function of the optical depth:
$$
g(\tau) = (1-\frac{3}{2} e^{-(\tau_0  + 2)}) cos \ \frac{\pi}{2}
(1 - \frac{\tau}{\tau_0} ) + \frac{3}{2} e^{-(\tau_0 + 2)} ,
\eqno{(33)}
$$
where, $\tau_0$ is the total Thomson optical depth of the CENBOL region and 
by construction $g(\tau_0)=1$. This is easily translated in radial 
coordinate for a typical flow model and used in the energy equation 30(d).

The general procedure of solving this set of simultaneous differential 
equations is provided in [25] and in [5] in detail. Although the 
flow deviates from a Keplerian disk to pass through a sonic point, 
and therefore the sonic point properties are to be obtained a
posteriori, it is best to assume the location of the sonic point
as well as the angular momentum at that point (or, alternatively
at the horizon) along with a suitable viscosity parameter. 
The solutions are integrated outward till they reach a 
Keplerian disk. This way the shock-free solutions are obtained. 
Most of the `shock-free' solutions which pass through the outer saddle
type sonic points do pass through shocks and then through the inner sonic
points on their way to black holes and neutron stars. (Careless
computations usually miss these solutions.) To search for 
solutions which include shocks, one has to incorporate Rankine-Hugoniot 
conditions.

The complete solutions from regions ${\cal E} \geq 0$
with and without shocks cannot join with a cold Keplerian flow
even when viscosity is added, since these flows are not bound.
If one writes the net energy (Bernoulli constant) as
$$
{\cal E} = \frac{1}{2} v^2 + \frac{1}{2}\frac{l^2}{x^2}+ 
\frac{1}{\gamma-1} a_s^2 - \frac{1}{2(x-1)}
\eqno{(34)}
$$
our arguments will be clearer. (Here, the rest mass energy in ${\cal E}$ 
has not been included. $a_s$ is the adiabatic sound speed). For a cold 
Keplerian disk, sound speed $a_s \sim 0$, $v\sim 0$, and $l_K = \frac{1}
{2}\frac{x^3}{(x-1)^2}$. At the junction point, where the advective disk 
meets the Keplerian disk, $l= l_K$ and ${\cal E} = \frac{(2-x)}{4(x-1)^2} <0 $ for
all $x>2$. Only when the disk is very hot ($1>a_s>>0$), or the flow is away from the 
equatorial plane (where potential energy is smaller than its value on the
equatorial plane) or matter coming out of cold disks and eventually 
heated up by, say, magnetic flares or dissipation, can have specific energy larger 
than $0$ and can join with the advective disk solutions. Note that these hot, 
energy conserving solutions are for strictly inviscid flow. The entire energy 
of the flow is advected to the black hole rendering the disk 
to be non-luminous. It is proposed that this may be the reason
why our galactic center is also faint in X-rays [6], although arguments
based on total luminosity is usually not a full proof. It is interesting  
that the same set of equations 30(a-d) shows a rich variety of time 
dependent behaviour. The implications would be discussed in the next Section.

Complete set of topologies of the viscous solutions are presented
in [25] for isothermal flows. They remain identical even when the
assumption of isothermality is dropped [5, 6]. For the sake of completeness
of the review we reproduce here some of the selected solutions of [25]
already presented in Chakrabarti [58]. Typical solutions 
are shown in Fig. 6(a-d) and the corresponding angular momentum ($\lambda$) 
distributions are shown in Fig. 6(e-h). Each solution is identified by
only three parameters, namely, the inner sonic point $r_{in}$,
the specific angular momentum at the sonic point $\lambda_{in}$ and the
constant viscosity parameter $\alpha$. The closed solutions of Fig. 4
open up in presence of viscosity. For low enough viscosity, shock
condition may still be satisfied as in Fig. 6a, but as $\alpha$
is increased (6b), $\lambda_{in}$ is reduced (6c), or $r_{in}$ is reduced,
the topologies change completely. The open solution passing through the
inner sonic point joins with a Keplerian disk at $r_{K}$. 
This change of topology triggered by the variation of viscosity may be 
considered the singlemost important development in the study of the 
accretion processes in the recent past. For a given cooling process (mainly governed 
by the accretion rate) $r_K$ strongly depends on viscosity: 
higher the viscosity, smaller is $r_K$. The paradoxical property
is primarily responsible for the observed nature of the novae outbursts [23]
as well as hard and soft states of black holes.
Similar situation occurs if parameters are taken from 
the region $I^*$ where only inner sonic point is present. 
The outer sonic point is also present for flows with positive specific energy, 
and thus, in principle, the solutions passing through the outer sonic point
may also join with a Keplerian disk. However, we suspect that in absence
of stable shock solutions, flows in 6(b-d) would produce unstable oscillatory
behaviour. The region between the Keplerian disk and the
black hole is basically freely falling, till close to the horizon
($x \sim l^2 $; note that angular momentum is 
nearly constant close to the black hole) where the centrifugal barrier is formed
and matter slows down, heats up and is puffed up just like
a constant angular momentum thick disk. 
Highly viscous Keplerian disk stays in the equatorial plane till $r_{K}$
and then becomes sub-Keplerian (a part of the flow may also become
super-Keplerian before becoming sub-Keplerian, [5]) as the flow enters through
the horizon. If the viscosity monotonically decreases with height, the 
flow would separate out of a Keplerian disk and form sub-Keplerian 
halo at a varying distance depending on the viscosity coefficient $\alpha (z)$.
Thus, typically a generalized accretion disk would have the shape as
shown in Fig. 3(c). The centrifugal barrier closer to the 
horizon may or may not be abrupt, depending on the parameters involved.
In either case, the flow density, temperature, velocity etc.
remain very similar as is shown in Fig. 7, where two solutions,
one with and the other without a shock are plotted. Thus the properties
of CENBOL is independent of whether a shock actually forms or not. 
For comparison, a high viscosity flow solution is also presented 
which deviates from a Keplerian disk closer to the black hole.

In passing, we may mention that apart from
Rankine-Hugoniot (non-dissipative) shocks, global solutions  also exist
where the shocks themselves dissipate a large chunk of the flow energy. 
An infinite number of such one parameter family of dissipative 
shock solutions are in the literature [59].
Shock solutions in various advective disk models have also been
obtained by several authors quite independently [49, 53-54, 60]
with similar properties. 

\section{Behaviour of Matter Around a Black Hole: Results of Numerical Experiments}

One of the most convincing ways to check 
if the inviscid and viscous solutions were stable or not
is to perform numerical simulations (with a reliable code, of course!). 
Recently, all possible fully time dependent behaviour of the advective disks 
have been found [50-51, 56, 60-63].
Shock solutions (from the region $SA$ 
in Fig. 4) were found to be stable and the results are in good agreement 
with theoretical predictions. Indeed, present development may 
be considered to be the best known way to test a code in spherical 
and cylindrical coordinates when shocks are present. Figure 8 shows the 
theoretical and numerical simulation results where the results from three 
completely different methods (smoothed particle hydrodynamics, 
total variation diminishing and explicit/implicit
code). This development is to be compared with the
poor matching of the solutions in early days of numerical simulations [64]
when both the theory as well as the code were not satisfactory.
Shocks also form in two dimensional (axisymmetric thick) flow
very near the predicted locations. When the solutions have one 
sonic point and shocks are not predicted (in regions O and I) 
shocks do not form (uppermost and lower most sets of curves). 
When the solutions have two sonic points but still shocks do not 
form (in region $NSA$ of Fig. 4), the shocks may form nevertheless, 
but they oscillate back and forth thereby changing
the size of the CENBOL [50]. In presence of cooling 
effects, shocks may oscillate even when stable shocks are
theoretically predicted  [65].
This typically happens when the cooling time scale roughly agrees with the 
infall time scale. The oscillating shock has the period comparable 
to the cooling time and is believed to explain the quasi-periodic
oscillations observed in the black hole candidates. The viscous
flows also show the similar oscillations [63].
We suspect that whenever accretion rates of 
a black hole change substantially (such as when a black hole 
changes its spectral state), the oscillations may be set in as a result of
competition among various time scales.  It is generally the case
that shocks are `always' formed whenever some angular momentum is present! They
may or may not be stable, i.e., they may be transient and propagate
away to a large distance, or, they may be oscillatory, or, they may be standing.
The exact behavior depends on the flow parameters.

One of the intriguing questions remained: Although advective solutions
produce sub-Keplerian flows from a Keplerian disk, are they really stable?
Chakrabarti et al. [62] provides a collection of numerical simulation
results including the formation of an advective disk from a 
Keplerian one far away from the black hole (and not just near the horizon
as in [47]). Figure 9a shows the ratio of the disk 
angular momentum to the Keplerian angular momentum in one of the simulations.
The shape is typical of such advective flows (See Fig. 10 of [20]) although
the transition from Keplerian to the advective disk is not very {\it smooth}. 
At a first sight, the reason seems to be due to the fact that the derivative 
$dl/dr$ in an advective flow is different from that in the Keplerian disk. 
One could find a smoother transition by adjusting viscosity and 
cooling effects at the transition. Figure 9b shows the deviation of 
angular momentum of a flow which included a standing shock. Apart from 
a mild kink in the distribution at the shock location, the flow is 
perfectly smooth, transonic and stable.

\section{Behaviour of Matter and Stars Around a Black Hole: Observations}

\subsection{From Spectral Properties}

Black holes are being fundamentally black, their proper identifications
must necessarily include quantification of very special spectral 
signatures of radiating matter entering in them. The inner boundary 
condition of the flow is unique and this 
automatically separates the true solution from a large number of
spurious solutions. Because of this, the spectral properties 
of the flow entering in a black hole should be different.
The problem lies in quantification of this special character. Here
we present a few observational results and how they may be readily 
understood using theoretical results presented in the earlier
Section. The advantage of this approach is that the
explanations are general (as they are straight from solutions of 
governing equations), and do not depend on any particular black 
hole candidate. In what follows, the accretion rates are 
expressed in units of the Eddington rate.

\subsubsection {Hard and Soft states and triggering of their transitions}

Galactic black holes are seen basically in two states. In soft states, 
more power is in soft X-rays and in hard states more power is in 
hard X-rays. (The extragalactic cases such state separation
is not obvious, since the observations are poorer, and transition
of states may take place in thousands of years. Some of the carefully observed
cases the spectral nature was found to be similar to those of the
galactic candidates, e.g., MRK841  [66].
Fig. 10(a-c) shows a comparison of the schematic spectra of
a black hole candidate both in hard and soft states and a neutron star candidate. 
Generally, neutron star spectra are composed of a multicolour 
black body component coming out of a Keplerian disk and a 
black body component coming out of the stellar boundary layer.
They always show soft bumps, unlike in a black hole in hard states, 
where the soft bump disappears. A neutron star does not show the 
weak power-law component which is really the hallmark of a black hole 
in soft states. The explanation of this apparently puzzling state
variation may be simple: the Keplerian and sub-Keplerian components redistribute
matter among themselves depending on viscosity of the flow which, at the
same time, also change the inner-edge of the Keplerian component. Sudden rise
in viscosity would bring more matter to the Keplerian component (with rate 
${\dot m}_d$) and bring the Keplerian edge closer to the black hole
(see, [47] for numerical simulation of this effect) and sudden fall of viscosity
would bring more matter to sub-Keplerian halo component (with rate ${\dot m}_h$) 
and takes away the Keplerian component away. Disk component ${\dot m}_d$ not only 
governs the soft X-ray intensity directly coming to the observer, it also 
provides soft photons to be inverse Comptonized by sub-Keplerian CENBOL 
electrons. The CENBOL (comprised of matter coming from ${\dot m}_d$ and 
${\dot m}_h$) will remain hot and emit power law (energy spectral index, 
$F_\nu \sim \nu^{-\alpha}$, $\alpha \sim 0.5-0.7$) hard X-rays only when
its intercepted soft photons from the Keplerian disk
(See Fig. 4) are insufficient, i.e., when ${\dot m}_d <<1$ to ${\dot m}_d
\sim 0.1$ or so, while ${\dot m}_h$ is much higher. For ${\dot m}_d \sim 
0.1-0.5$ (with ${\dot m}_h \sim 1$), CENBOL cools catastrophically and no
power law is seen (this is sometimes called a high state). 
With somewhat larger ${\dot m}_d$,  the power law due to the bulk motion of 
electrons [20, 67-68]
is back at around $\alpha \sim 1.5$ (this is sometimes called 
a very high state). Figure 11 (taken from [21])
shows a typical hard to soft state transition as ${\dot m}_d$
is increased. Here, power $E F(E)$ is plotted against the energy $E$
of the emitted photons. The dashed curve drawn for ${\dot m}_d=1.0$ 
includes the convergent flow behaviour of the inner part of CENBOL.
Details of the solutions are in [20, 21, 23].
Such hard/soft transitions are regularly seen in black hole
candidates [69-73].

\subsubsection{Constancy of Slopes in Hard and Soft States}

Spectra of the advective disk solutions show a remarkable
property: the slope $\alpha \sim const $ in hard states even when
${\dot m}_d$ is increased by a factor of a thousand. The degree
of constancy is increased [21] if one assumes that as the
viscosity changes, the matter is actually redistributed between 
the Keplerian and sub-Keplerian halo components 
rather than assuming that both the components are completely independent. 
This constancy of slopes is regularly seen [20, 23, 72, 74-75].
Particularly important is the weak power law in the soft state 
as this is not observed in neutron star candidates. CENBOL 
around neutron stars may also cool down to produce soft state for 
the same reason. However, they can go up to high state and not up to 
`very high' state where the weak power law due to convergent flow is seen. 
In bulk motion Comptonization bulk momentum of the quasi-freely
falling electrons (outside the horizon) are transported to the soft photons
(Doppler Effect -- [20, 67-68]).
This effect becomes important (compared to the thermal Comptonization) when the
electrons themselves are cool (less than a few keV).
In neutron stars, electrons slow down on the hard surface due to radiation
forces opposite to gravity acting on them, and therefore the bulk
momentum transfer is negligible. Thus black holes could be identified by 
spectral signatures alone provided they are seen in soft states [4, 20].

\subsubsection{Variation of Inner Edge of the Keplerian Component}

This is a trivial property of the advective disks (see, [5, 20]).
As viscosity is increased, the location $r_{K}$ where the disk
deviates from Keplerian is {\it generally} decreased if other two 
parameters ($r_{in}$ and $l_{in}$) are held fixed. Thus,
in hard states, not only ${\dot m}_d$ is smaller, the $r_K$ is also larger.
As the viscosity increases, $r_{K}$ becomes smaller in viscous
time scale, at the same time more matter is added to the Keplerian
component. This behaviour is also seen in black hole candidates [76].
In advection dominated models of Narayan \& Yi [27] such variations are achieved by 
evaporation of the disks by unknown fundamental physics.

\subsubsection{Rise and Fall of X-ray Novae}

X-rays novae (e.g., A0620-00, GS2000+25, GS1124-68, V404 Cygni 
etc.)  produce bursts of intense X-rays which decay with time 
(decay time is typically 30d). This phenomenon may be repeated 
every tens to hundreds of years. While in persistent black 
hole candidates (such as Cyg X-1, LMC X-1, LMC X-3) Keplerian 
and sub-Keplerian matter may partially redistribute to change 
states (see Section 3 above), in X-ray novae candidates the net 
mass accretion rate may indeed decrease with time after the outburst, even 
if some redistribution may actually take place. First qualitative 
explanation of the change of states in X-ray novae in terms of the 
advective disk model was put forward by ETC96. The biggest
advantage of the advective solution is that it automatically 
moves the inner edge of the Keplerian disk as viscosity is varied.
Similar to the dwarf novae outbursts, where the Keplerian disk 
instability is triggered far away (e.g., [77])
here also the instability may develop and cause the viscosity
to increase, and the resulting Keplerian disk with higher
accretion rate moves forward. In [21], several such spectral 
evolutions have been presented. In Fig. 12, one such case 
is shown, where the increase in viscosity is used to cause 
the decrease in $r_K$ from $9000$ to $10$, keeping ${\dot m}_h =1$
and ${\dot m}_d=0.01$. As the inner edge goes from $9000$ 
to $5000$, the optical (around tens of eV) peaks first, which 
is followed by hard X-rays (at around hundreds of keV) till 
$r_K$ reaches about several hundred Schwarzschild radii. 
After that the hard X-ray subsides and soft X-ray intensifies. 
The optical precursor of an X-ray nova GRO J1655-40  
have been seen recently [78].

\subsubsection{Quiescent States of X-Ray Novae Candidates}

After years of X-ray bursts the novae becomes very faint
and hardly detectable in X-rays. This is called the
Quiescent states of the black holes. This property is
is in built in Advective disk models. As already 
demonstrated [5, 20],
$r_{K}$ recedes from the black 
holes as viscosity is decreased. With the decrease of
viscosity, less matter goes to the Keplerian component [47],
i.e., ${\dot m}_d$ goes down. Since the 
inner edge of the Keplerian disk does not go all the way to the
last stable orbit, optical radiation is weaker in comparison
with what it would have been predicted by a Shakura-Sunyaev [11]
model (see, plots for $r_K=9000$ and $8000$ for such spectral 
behaviour in Fig. 12 above). This behaviour is seen in V404 Cyg [79]
and A0620-00 [80]. 
The deviated component from the Keplerian disk almost resembles a constant 
energy rotating flow described in detail in [3]. It is also 
possible that our own galactic center may have this low viscosity, 
low accretion rate with almost zero emission efficiency [3]
global advective disks as mentioned in [6].

Recently, a so-called advection dominated model has been used to
fit these states [81].
Though it is supposed
to be an off-shoot of the advective disk solutions for
low optical depth limits, this is not a self-consistent model.
In this model highly viscous ($\alpha \sim 0.1-0.5$) quasi-spherical 
flow resulted from Keplerian disk evaporation (which is also in 
equipartition with magnetic field at all radii!) was used. 
The data contained typically `one point' in the hard X-ray region
and fits are poor.  On the contrary, advective disk solution [3, 5, 20]
does not require such evaporation, and the advecting ion torus 
of low mass accretion rate comes most naturally out of 
the governing equations only for very low viscosity case. 
The deviation from a Keplerian takes place several thousand 
Schwarzschild radii. In high viscosity case angular momentum transport
rate becomes so high that the flow deviates from a Keplerian disk
almost immediately from the inner sonic point.
Advecting disks produce the quiescent state like spectra [20-21] 
without making any further unwarranted approximations, regarding magnetic 
equipartition, etc. However, so far, advective disk solutions,
theoretically complete as they are, have not made any 
serious attempt to fit observational data yet.

\subsubsection{Quasi-Periodic Oscillations of X-rays}

As mentioned in Section 2.4, in some large region of the parameter space
the solutions of the governing equations 30(a-d) are inherently
time-dependent. Just as a pendulum inherently oscillates, the
physical quantities of the advective disks also show oscillations of the
CENBOL region for some range in parameter space.
This oscillation is triggered by competitions 
among various time scales (such as infall time scale, cooling 
time scales by different processes). Thus, even if black holes
do not have hard surfaces, quasi-periodic oscillations could
be produced. Although any number of physical processes such as
acoustic oscillations [82],
disko-seismology [83], 
trapped oscillations [84],
could produce such oscillation frequencies, modulation of $10-100$ 
per cent or above cannot be achieved without bringing in the dynamical 
participation of the hard X-ray emitting region, namely, the CENBOL.
By expanding back and forth (and puffing up and collapsing, alternatively)
CENBOL intercepts variable amount of soft photons
and reprocesses them. Some of the typical observational
results are presented in [85-87].
Recently more complex behaviour has been seen in GRS 1915+105 
[88-89],
which may be understood by considering  several cooling 
mechanisms simultaneously. Particularly interesting is the observations [88]
that the rise time is slow (10-15s) but the decay time is rapid
(2-3s). This is typically interpreted as the signature of rapid swallowing of matter into a 
black hole. QPOs in neutron stars do not show this property.
Some chaotic behaviour of $r_K$ under non-linear feed back 
mechanism cannot be ruled out either. 

\subsubsection{Nature of the Iron Line}

Resonance lines of iron have been seen in several black hole candidates
[90, 91].
Usually one of the two observed wings is found to be stretched compared to the
other and it is explained  to be due to the combination of the
Doppler shift and the gravitational red-shift. Generally, it is
difficult to explain very large equivalent width of the lines in this
models.  This problem can be circumvented if the lines are
assumed to be coming from outflowing winds. The stretched wing
would then be due down-scattered emission lines [20]. The idea
of line emissions from the winds is finding supports by other workers
as well [91-93].

\subsection{From Motion of Stars}

Measurements of the stellar rotation and velocity dispersion profiles
close to a galactic nucleus can provide the mass of the central body.
For a spherical distribution of stars, for instance, virial mass within
radius is obtained as $M(R) = \frac{\beta \sigma_v^2 R}{G}$, with $\beta$ 
varying from $2$ to $3$. Using the dispersion and rotation, Kormendy 
[94] computed the mass of the central nucleus of M31 to be a few $\times 
10^7 M_\odot$. Its companion M32 was probably the first galaxy with a convincing
evidence of a massive black hole of about $3 \times 10^6 M_\odot$ 
(see, [95] and references therein). Our own 
galactic center is now being studied extensively by Genzel group [96].
The projected stellar velocity
dispersion increases significantly from $55$km s$^{-1}$ 
at around $5$pc, to $180$ km s$^{-1}$ at around $0.1$pc. The
estimated mass is $2.5-3.2 \times 10^6 M_\odot$. The corresponding
mass density is more than $6.5 \times 10^9 M_\odot$pc$^{-3}$. This is one 
of the highest observed concentration of matter that has been detected so far.

\subsection{From Mass Functions}

Some argue that measurement 
of the central mass is the most definite way to identify a back hole.
In a binary system, the mass of the compact primary $m_1$ is estimated from the
observed mass function $f(m)$ and the inclination of the orbital plane $i$ (e.g., [1]):

$$
M_1 = \frac{f(m) (1+ q)^2}{sin^3 i}
\eqno{(35)}
$$

Since $M_1$ is the minimum mass of the compact object, if its
value is larger than $3 M_\odot$, the object is most probably a black hole.
satisfy these criteria and  the compact
components are possibly a black holes. The most well studied suspected black hole
component Cyg X-1 (1956+350) has mass function of only $0.24 M_\odot$ 
and therefore it fails this criterion. However, spectral signatures, 
such as the transition of states  [73, 97]
and weak hard tail in the soft state positively identify it to be a black hole.

\subsection{From Doppler Shifts}

As matter rotates around a black hole, the line emissions
produce well known double horned pattern seen in
disks around compact stars. The origin of this spilt
lies in the Doppler effect. Here the frequency of
matter increases as matter comes towards an observer and
it decreases as matter goes away from an observer. The
change of frequency is roughly $\delta \nu /\nu \sim v/c$,
where $v$ is the rotational velocity projected along the
line of sight. In a Keplerian disk, $v_K =\sqrt(GM/x)$.
Thus knowing where the line is actually emitted, one could
compute the mass of the central object. Through
water mega-maser experiments one has been able to detect this
strictly Keplerian motion in NGC4258 [98]
and the estimated mass of the central black hole is $4 \times 10^7 M_\odot$.
Observation of the disk around M87 was made using Hubble space telescope 
and the Doppler shift is clearly observed in flows within 
the disk. Estimate of the mass of the black hole assuming a Keplerian disk is 
around $(2.0 \pm 0.9) \times 10^9 M_\odot$ [99].
However, the interpretation of the Doppler shift varies. It is doubtful
whether the lines could be emitted at large distances 
($\sim 20-80$ thousand Schwarzschild radii from the center) 
from a Keplerian disk, as the disk would be too cold 
(unless the disk is strongly warped as NGC4258, and intercepts its
own X-rays emitted at the center of the galaxy; though
M87 is not considered to be particularly bright in X-rays).
An alternate explanation would be that the line emission farther away
from the center is (spiral-)shock excited. However, flows with spiral 
shocks (any shocks for that matter) must be at least partially sub-Keplerian
(even if pre-shock flow is Keplerian, post-shock flow must be sub-Keplerian
for any shock strength)
and therefore the estimated mass from the same shift in
frequency must be higher. This suggestion produces
a mass of much higher value  [100]
-- $4\pm 0.2 \times 10^9 M_\odot$ within $10^4 R_g$, i.e., about $3.5$pc.
This is the first time where, using the suggestion of
even a partially sub-Keplerian disk in galaxies,
higher central black hole mass estimate has been made.
It was farther noted that the spiral shock should not be extended
below $3.5$pc, otherwise, the width of the lines would be higher. Recently, Macchetto et al. 
[101] did find this deviation from Keplerian
in the outer edge of the disk. However, central of this disk is
very close to Keplerian. Using high resolution ($0.09"$) observation of Hubble
Space Telescope in this region they also found the mass to be
in this high range ($3.2 \pm 0.9 \times 10^9 M_\odot$).
Thus, results from Keplerian and sub-Keplerian regions independently
give rise to the same high black hole mass.
Another object whose mass has been obtained recently using Doppler
effect is M84. Long-slit spectrophotometry is used in mapping the velocity
profile across the  disk and assuming a Keplerian disk, the
mass is found to be $1.5 \times 10^9 M_\odot$ [102].

\subsection{From Reverberation Mapping}

Generally, active galactic nuclei also show line emissions along with 
continuum emissions. These lines are believed to be emitted from
rapidly moving clouds on either sides of an accretion disk. Measurement 
of the motion of the cloud from Doppler shift and the distances 
of the cloud from reverberation mapping [103]
method can give an estimate of the mass of the central object. In this method,
the time lag between certain variation in the continuum spectra and the
line emission is used to measure the distance of the broad line emitters.
Masses of a few active galaxies have been measured this way:
NGC 5548 ($8.8 \times 10^7 M_\odot$), NGC 3227 ($3.8 \times 10^7 M_\odot$) etc.

\subsection{Comparative Studies of Detection Mechanisms}

There are several ways such as those using motion of stars, 
Doppler effects, mass function, spectral features, gravitational 
wave (which has not been detected yet) etc., for the identification 
of a black hole. Except for the spectral feature study 
(that too, for instance, in the very soft state where the 
weak power-law region is observed) and the gravitational wave 
signals, where the effect of the horizon could be seen, 
all the other criteria cannot really distinguish a black hole 
(with a horizon) from another hypothetical solution of Einstein equation 
which is massive and `somehow' compact. In several cases, particularly,
in the cases of active galaxies, one may have to be satisfied
with the more indirect methods such as Doppler shifts, reverberation
mapping etc. since transition of states from hard to soft, or vice versa,
might take a very {\it long} time.

\section{Signatures of Advective Flows in Other Branches of Astrophysics}

If the flow is highly advective around a black hole, 
and at the same time shock structures and CENBOL form,
as the solutions seem to indicate, there are a large number of
spin-offs which should also be observable. For instance, jets and
outflows are known in many systems. Are they related to 
the degree by which the flow is sub-Keplerian or the
formation of CENBOL? If so, can the outflow rate be estimated
from such considerations? Similarly, since the temperature of the
advective disks is very high, could it cause a significant nuclear
burning in the disks? Could the non-Keplerian disk influence
gravitational wave emission from a coalescing companion
which happens to interact with the disk as well? There answers
seem to be `yes' to all, and major developments in these directions
have been done very recently. We discuss them here very briefly.

\subsection{Physics of Jets: Estimation of the Outflow Rate From an 
Advective Flow}

Outflows are common in many astrophysical systems which 
contain black holes and neutron stars. Difference between stellar
outflows and outflows from these systems is that the outflows in these
systems have to form out of the inflowing material only, whereas
in stars outflows are `extensions' of the staller atmosphere.
Although a black hole does not have a hard surface, the centrifugal barrier
behaves like one, and therefore mass loss associated with it could be 
computed. Of course, a shock surface around a black hole 
need not be only centrifugally supported. Chang \& Ostriker
[104] produced shocks assuming considerable pre-heating of the
incoming flow and Kazanas \& Elison [105] suggested shock formation
which are supported by pair plasma pressure. As long as such a 
region where some compression (other than geometric, with shock or no shock)
is formed in the inflow, the jet formation take place. It is to be noted that
production of jets have always been found to be favourable when the
disk itself is sub-Keplerian [106].

Figure 13 shows schematically the `black box' where outflows are
generated. Assuming this configuration, mass loss is estimated 
very easily [107]. The procedure involves in first computing the
CENBOL temperature $T_s$ from the incoming flow using steady
shock condition and then use the same procedure as used on stellar 
surface, namely, mass loss from this surface using transonic
flow condition. The ratio of outflow rate to the inflow rate in terms of the 
compression ratio $R$ of gas turns out to be,
$$
R_{\dot m}=\frac{{\dot M}_{out}} {{\dot M}_{in}}=\frac{\Theta_{out}}
{\Theta_{in}} \frac{R}{4} exp (-f) f_0^{3/2}
\eqno{(40)}
$$
where, $ f= f_0 - \frac{3}{2}$ and  $f_0=(2n+1)R/(2n)$
[$n=(\gamma -1 )^{-1}$ is the polytropic constant.].
Notice that this simple result does not depend on the
location of the sonic point or shock (namely the size of the
dense cloud, and physical process which produces it) 
or the outward force causing the mass loss. It
is a function of compression ratio $R$ for a given geometry.
In a relativistic inflow $n=3, \ \gamma=4/3$ and $R=7$ and the ratio
of inflow and outflow becomes,
$$
R_{\dot m}=0.052 \frac{\Theta_{out}}{\Theta_{in}}
\eqno{(41a)}
$$
and for inflow of an ionized gas $n=3/2, \ \gamma=5/3$ and $R=4$,
and the ratio in this case becomes,
$$
R_{\dot m}=0.266 \frac{\Theta_{out}}{\Theta_{in}}
\eqno{(41b)}
$$
Outflows are usually concentrated near the axis, while the inflow is near
the equatorial plane. Assuming a half angle of $10^o$ in each case, we obtain,
$$
\Theta_{in}= \frac {2 \pi^2}{9}; \ \ \ \ \ \Theta_{out}= \frac {\pi^3}{162}
\eqno{(42)}
$$
and 
$$
\frac{\Theta_{out}}{\Theta_{in}} =\frac{\pi}{36} .
\eqno{(43)}
$$
The ratios for $\gamma=4/3$ and $\gamma=5/3$ are then
$$
R_{\dot m}= 0.0045\ \ \  {\rm and} \ \ \ R_{\dot m}=0.023
\eqno{(44)}
$$
respectively. This is to be compared with the rate $R_{\dot m}=0.004$
found in radiation dominated flow [108].
A more exact computation of the mass loss rate could be done
using exact transonic solutions for the inflow and outflow [109].
Fig. 14, shows the actual
solution of isothermal winds coming out of adiabatic accretion
for a rotating flow with $\lambda = 1.89$ and ${\cal E}= 0.0038$.
Three outflowing dashed curves correspond to three accretion rates 
(decreasing monotonically from the  upper curve  to the lower one).
In this calculation, the outflowing region is 
assumed to be between the centrifugal barrier and the funnel wall, and the
Comptonization (Chakrabarti \& Titarchuk, 1995) has been
qualitatively incorporated to compute the CENBOL temperature as a function of the
accretion rate. The general conclusion in this case is that
the percentage of the outflow rate is non-linearly dependent on the 
inflow rate even for a constant angular momentum flow, since the
temperature of the CENBOL which drives the flow is a steep 
function of ${\dot M}_{in}$. For lower accretion rate, the Comptonization
is inefficient and the ratio of outflow to inflow increases.

\subsection{Nuclear astrophysics: Nucleosynthesis in advective disks 
around black holes}

Chakrabarti [110] and Chakrabarti, Jin \& Arnett [111], first pointed 
out that a considerable nucleosynthesis could take place during the 
infall and heavier elements may be produced inside the thick disk,
a fraction of which could be ejected out through bipolar outflows 
and jets [112]. Although the disk model used was 
very preliminary (basically then fashionable purely rotating
thick accretion disk with a slow infall) 
the conclusions were firm and were verified by a 
large number of independent workers [113-114]
using other disk models. 

In the decade since these works were initiated,
the self-consistent advective disk model has been developed.
This motivates one to look into the nucleosynthesis problem
once more, specially when the shocks and CENBOL regions are
also included in the computation. Fig. 15 shows the variation of the
abundance as the matter enters the advective disk regime [115].
Here a $10 M_\odot$ central object and a (total: Keplerian and 
sub-Kepelrian combined) mass 
accretion rate of $1{\dot M}_{Edd}$ are used.
Comptonization is roughly incorporated by reducing the
proton temperature by a factor of $30$ as is seen in [20]
in this case. The cooling factor $f=0.5$ and viscosity 
$\alpha_\Pi=0.05$ were used which gave $x_K=498 r_g$. The 
shock is formed at $x_s=13.9 r_g$ (see, Fig. 4 for the 
full solutions). 
The dotted curves are drawn when only the supersonic branch through outer 
sonic point is used, while the solid curves are drawn when the solution
takes more stable branch through the shock and finally through
the inner sonic point. At the shock, the sudden 
rise in temperature as well as higher residence time in the post-shock 
flow causes the abundance to change abruptly although the final product
at the horizon remains very similar. Only the elements with abundance
above $10^{-4}$ at any stage during the inflow are indicated, although
a total of 255 isotopes (from neutron, proton, helium to germanium)
were used in the network. This is a more `dramatic' case. Generally,
when accretion rate is high, the advective region is small, so the
effect is much weaker. When accretion rate is low, advective region
is larger only when the viscosity is low also. But density being smaller,
the effect (i.e., change in abundance) is also weaker.

In the high accretion rate case, heavy elements are produced
and light elements are destroyed,
while in the low accretion rate 
case, one may imagine that spallation reaction becomes important 
to produce some excess $Li^7$ [116-117]. 
However, detailed computation shows that before the spallation
could take place, entire $He^4$ may be destroyed due to photo
disintegration [118].
This is because, even though the accretion flow is of low rate
and photons emitted by itself could be low, photons may be supplied
externally by the cooler Keplerian disk. These cooler
photons are energized by inverse Comptonization and
they can then participate in photo-disintegration process.
Thus, the production of Li$^7$ or D does not seem to be possible
under any circumstances in the black hole accretion.
One interesting physical process make the nucleosynthesis study
much more important: the production of neutron tori [119].
Neutrons produced in the advective disk due to 
photo-dissociation produce a neutron tori in the advective region which, 
mixing with fresh incoming matter may produce neutron rich isotopes 
of the galaxies. 

The changes in abundance in the CENBOL region is important, since the wind
may be produced from this region at a rate ${\dot M}_{out}$ (see Section 5.1).
Part of wind could be intercepted by the companion star and it is likely that
these new elements may be detected in the stellar atmosphere.
In this context, recent observations of high Lithium abundance
in K-star companions of black holes [120]
may be significant,
although this may be entirely due to magnetic flaring on the stars [121].
and quasars, heavy elements produced in the disk may supply metalicity in the galaxies.

Jin, Arnett and Chakrabarti [122] originally concluded that the
effect of nucleosynthesis is important only for very low
viscosities. This is because they focused on cooler radiation dominated 
disks where higher residence time was required.
Using present advective disks, even for $\alpha \sim 0.1-0.4$, 
nucleosynthesis seems to be important for stellar black holes [115].
For supermassive black holes 
the effects remain weaker since the density of the disk becomes
much lower for a comparable non-dimensional accretion rate.

\subsection{Gravity Wave Astronomy: Effects on Gravitational Wave Emission}

Traditionally, coalescence of two compact bodies is studied in the
absence of accretion disks. This may be justified where both the 
components in a binary are compact, such as two neutron stars and two 
stellar mass black holes. When a supermassive black hole at the galactic 
center is surrounded by an advective disk through which a compact
companion star gradually moves in along an instantaneously
Keplerian orbit, not only does the angular momentum of the 
companion is lost due to gravitational wave emission, but some angular
momentum is also changed through the interaction of the non-Keplerian
part of the disk with the companion. For instance, in the 
super-Keplerian region of the disk, 
the companion will {\it gain} angular momentum due to accretion
from disk material, while in the sub-Keplerian region the companion
would lose angular momentum due to accretion of negative angular
momentum. In either case, the wave pattern of the emerging
gravitational wave would be affected. Through detailed computation
it was  shown that the disk effect could be up to $7-10$ percent of the main
effect [123].
Similar deviations from standard
template is also possible when self-gravitating disk is present,
since Keplerian angular momentum distribution of such disks are
completely different [124].

Figure 16, shows the effect of the presence of an accretion disk
on the gravity wave pattern in a binary black hole system
consisting of two black holes with mass $10^8 M_\odot$ and $10^6 M_\odot$.
The solution for the disk quantities is obtained from the
equations 30(a-d) for parameters $\gamma=5/3$, $\alpha= 0.02$, $f=0.0$,
$x_{in}=2.3$, and $l_{in}=1.7$. Accretion rate ${\dot M}= 1000 
{\dot M}_{Edd}$ is chosen for enhancing the effect. When ${\dot M}$
is reduced, as is appropriate for an advective disk, the effect 
is proportionately smaller. In Fig. 16a, the radial distance of the companion
as function of time is compared (solid curve is with the disk, and 
the dashed curve is without the disk), while in Fig. 16b, the `chirp' 
profiles as functions of real time are compared (only last few 
Schwarzschild radii are shown). 
When the sub-Keplerian disk is included, the companion falls
more rapidly due to enhanced loss of angular momentum. 

When viscosity is large the inner edge of the Keplerian component is closer
to the black hole and the soft state is achieved for higher 
accretion rates. For the same set of parameters, the 
effect on gravitational wave emission is maximum, since close to the
black hole the flow would still be advective and sub-Keplerian. 
In the hard states, the viscosity and accretion rate is low and the inner 
edge of the Keplerian component is farther out. In this case, the disk
will have a very little effect on the gravity wave emission. Thus, for the
first time, results from electromagnetic wave and the gravitational wave
could be combined to obtain a better understanding of the system parameters.
Detailed computations of the templates are in progress and would be
reported elsewhere.

\section{Concluding Remarks}

That black holes, which represent the end product of massive stars and 
star clusters, must exist somewhere in this universe is beyond any doubt. 
The issues discussed in this review were: whether they are in principle 
detectable, how to detect them and whether they have been detected. It seems 
that a few cases at least they {\it have been detected}. If the observations 
of Genzel et al. [96, 125] is correct, then the mass of the central $0.1pc$ 
region of our galaxy would be $2.5-3.2 \times 10^6 M_\odot$ and the 
corresponding mass density would be $6.5 \times 10^9 M_\odot /pc^3$, 
the highest measured concentration so far. The water mega-maser measurement 
of the nucleus of NGC4258 within $0.1pc$ has the central mass of
$4 \times 10^7 M_\odot$ and corresponding mass density is $6.5 
\times 10^9 M_\odot /pc^3$. The central mass of M87 from the estimation of
Keplerian and non-Keplerian components is $ \sim 4 \times 10^9 M_\odot$ and the
corresponding mass density is $2.0 \times 10^7 M_\odot /pc^3$. Although, 
Cyg X-1 is the most studied black hole candidate so far, its mass function is
very low. Its confirmation as a black hole comes from its spectral features,
especially the weak power-law slope of the bulk motion Comptonization in its
soft state. The only candidates with mass function higher than, say, 
$3 M_\odot$, are $GRS1124-683$, $GRO J1655-40$, $H 1705-250$, $GS 2000+25$ 
and $GS2023+338$ and are possible stellar mass black holes. With 
the improvements of the future observational techniques, one needs to 
focus on more detailed predictions of the advective disks, 
such as variation of the solution topology
with specific energy, or equivalently, accretion rate. With the
emergence of gravitational wave astronomy, the wave signals
from galactic centers should be detectable. The proposal
presented in Section (5.3) would for the first time correlate the
distortions of the gravitational wave signals with those
from the spectral signatures. Together they would not only
verify black holes, they may also become the strongest test
of general relativity to date.

\newpage

\centerline {Reference}

\bibliographystyle{plain}
{}


\newpage 
\centerline{Figure Captions}

Fig. 1a: Potential felt by photons in Schwarzschild space-time. Photons
with impact parameter $b < 3 \sqrt{3}$ would be swallowed by the black hole,
while those with $b >\gsim 3\sqrt{3}$ escape to infinity.

Fig. 1b: Photon trajectories with near-critical impact parameter 
$b=5.19615242$ at different radial distance $x$ (measured in units of
$GM_{BH}/c^2$). At each radial
distance the conical volume is drawn with semi angle $\psi= sin^{-1} 
[3\sqrt{3}/x \sqrt(1-2/x)]$. The shaded region is the `cone of 
avoidance' and the unshaded region is the `cone of emergence'.

Fig. 2: Potential barrier of a Schwarzschild black hole as felt by incoming 
test particle with angular momenta $l=0$, $2$, $2\sqrt{3}$, $3$, $4$, and $5$
(from bottom to top; in units of $GM_{BH}/c$). The dashed curve is the Newtonian
potential drawn for $l=2$ for comparison. Marginally bound and marginally 
stable orbit locations at $4$ and $6$ respectively (in units of $GM_{BH}/c^2$)
are indicated.

Fig. 3: Evolution of accretion flow model that takes place every twenty
years or so. Bondi flow (a), Keplerian disk (b) and the generalized 
advective disk (c) are shown. Beside each model typical spectrum is also shown
in $\nu log(F_\nu)$ vs. $log(\nu)$ scale. In (c) variation of the spectra
with viscosity (or, equivalently, relative accretion rates in the Keplerian and 
sub-Keplerian components is shown).

Fig. 4: Classification of the parameter space (central box)
in the energy-angular momentum plane in terms of various topology of the 
black hole accretion. Eight surrounding boxes show the solutions from 
each of the independent regions of the parameter space. Each small 
box shows Mach number $M$ against the logarithmic radial distance $r$
(measured in units of $2GM_{BH}/c^2$). 
Contours are of constant entropy accretion rate
${\dot{\cal M}}$. Similar classification is possible for all adiabatic index
$\gamma <1.5$. For $\gamma >1.5$, only the inner sonic point is possible
other than an unphysical `O' type point [5].  
See text for details.

Fig. 5: Variation of shock locations  with energy in accretion and winds for various
specific angular momenta $l$ (marked on curves). $a=0.5$ is chosen.
Segments marked $a_3$ and $w_2$ (solid curves) represent stable
shocks in accretion and winds respectively. Other two segments
($a_2$ and $w_3$) represent formal shock locations which are unstable.

\noindent Fig. 6: Mach number variation (a-d) and angular
momentum distribution (e-h) of an isothermal viscous transonic flow.
Only the topology (a) allows a shock formation in the steady flow. Transition to
open (no-shock) topology is initiate by higher viscosity 
($\alpha$) or lower angular momentum ($\lambda_{in}$) or inner sonic point 
location ($r_{in}$ measured in units of $2GM_{BH}/c^2$). These latter types produce nonsteady shocks [63].
In (e-h), flow angular momentum (solid) is compared with Keplerian angular momentum (dotted). 
The location from where the angular momentum is deviated varies with the 
three parameters. Figure is taken from Chakrabarti, [125].

\noindent Fig. 7: Ratios $v_x/v_\phi$ (solid) and densities (dashed) 
of three illustrative solutions of the advective flows [5]. Note that the 
centrifugal barrier close to the hole makes all the three solutions to
behave in the similar way in the region $2\lsim x \lsim 10-20$, emission from which 
strongly determines the spectral properties of the black hole.
In a strongly shocked flow the variations in densities and velocities
occur in a shorter length scale while in a weakly shocked 
or shock-free flows the variations occur in an extended region. 

\noindent Fig. 8:  Comparison of theoretical (solid) and numerical results in a 
one-dimensional accretion flow which may or may not allow a standing shock. Mach
number (y-axis) is plotted against the radial distance (in units of $R_g2GM_{BH}/c^2$).
The long and short dashed curves are the results of the TVD and SPH simulations
respectively while very long dashed curve is using explicit/implicit code.
The curves marked `O' and `I' are for transonic flows which pass through the
outer and the inner sonic points respectively. They are also reproduced perfectly 
with numerical simulations [62]. 

\noindent Fig. 9a: Ratio of disk angular momentum to the Keplerian angular
momentum in a typical time dependent simulation in an advective disk. The
plot is made after the steady state is reached. Note the
deviation Keplerian disk at around $R\sim 30$. The flow becomes super-Keplerian
close to the hole before becoming sub-Keplerian as it plunges in [62].

\noindent Fig. 9b: Numerical simulation result of a typical advective disk which forms
a standing shock after deviating from a Keplerian disk. While the
flow remains generally sub-Keplerian, a kink in the distribution at the
shock is produced, which, however, remained stable throughout the
simulation  [62]. 

\noindent Fig. 10(a-c): Comparison of the schematic spectra $F_\nu$ vs. $\nu$ of
a neutron star candidate  (a: left panel) and a black hole candidate both in soft 
(b: middle panel) and hard (a: right panel) states.
Neutron star spectra are composed of a multicolour black body component coming 
out of a Keplerian disk and a black body component coming out of the stellar envelop. 
In soft states a black hole shows a weak power-law component of a quasi-constant slope
($-1.5$). In hard states the power law component is stronger and has a slope of $\sim -0.5$.

\noindent Fig. 11: Spectral evolution of an accretion disk with a strong shock
at $X_s=10$ around a black hole of mass $3.6M_\odot$. The sub-Keplerian halo rate is
${\dot m}_h =1$ and the Keplerian rates are marked on the curves. The dotted curve is
drawn to include the effect of bulk motion Comptonization when ${\dot m}_d=1$ [21].

\noindent Fig. 12:  Typical spectral evolution of an X-ray nova. $x_K$ (marked on
the curves) could be very far away as in a low accretion rate, low viscosity
disk. We chose: ${\dot m}_h=1.0$ and ${\dot m}_d=0.01$. Initially, at the onset 
of an outburst, the optical intensity goes up as $x_K$ is decreased. Subsequently,
the hard X-ray goes up first and then the soft X-ray is intensified. The
$x_K=9000$ and $8000$ solutions resemble Novae spectra in quiescence [21].

\noindent Fig. 13: Schematic diagram of the incoming and outgoing flows around a black hole.
It is suggested that the CENBOL actually behaves like a stellar surface and causes the
mass loss exactly in the same way the stars lose mass [107].

\noindent Fig. 14: Actual solution  (Mach number along y-axis and log(r) along x-axis)
of isothermal winds (dashed curves) arising of adiabatic accretion (solid) for a rotating flow with $\lambda = 1.89$ 
and ${\cal E}= 0.0038$ in presence of a centrifugal barrier at $18$. Arrows indicate the
direction of the outflow. Outflow rate is increased from bottom to top outgoing curves.

\noindent Fig. 15: Variation of the abundance as matter enters the 
advective disk regime. Here $10 M_\odot$ central object and a mass 
accretion rate (sum of Keplerian and sub-Keplerian component) 
of $1{\dot M}_{Edd}$ is used. The cooling factor $f=0.5$ and viscosity 
$\alpha_\Pi=0.05$ were used which gave $x_K=498 r_g$. The 
shock is formed at $x_s=13.9 r_g$. The dotted curves are drawn when 
only the supersonic branch through outer sonic 
point is used, while the solid curves are drawn when the solution
passes through both the sonic points and a shock.

Figure 16: Effect of the presence of an accretion disk
on the gravity wave pattern in a binary black hole system
consisting of two black holes with mass $10^8M_\odot$ and $10^6 M_\odot$
See text for the parameters used.
In (a), the radial distance of the companion as a function of time are compared (dashed
curve is with the disk, and the solid curve is without the disk), 
while in (b), the `chirp' profile as function of 
real time are compared (profiles in the last few Schwarzschild radii are shown).
When the sub-Keplerian disk is included, the companion falls
more rapidly due to enhanced loss of angular momentum.

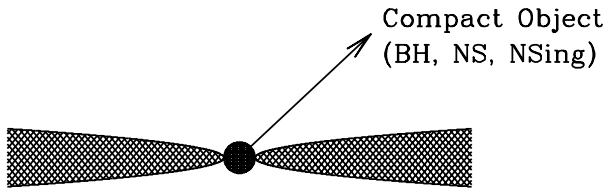
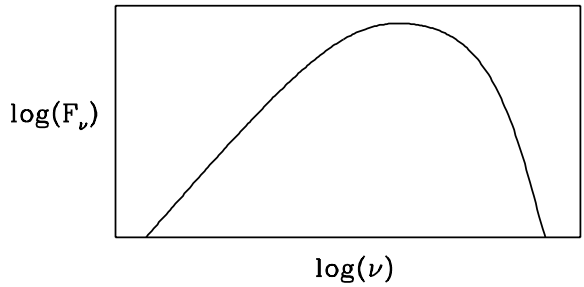
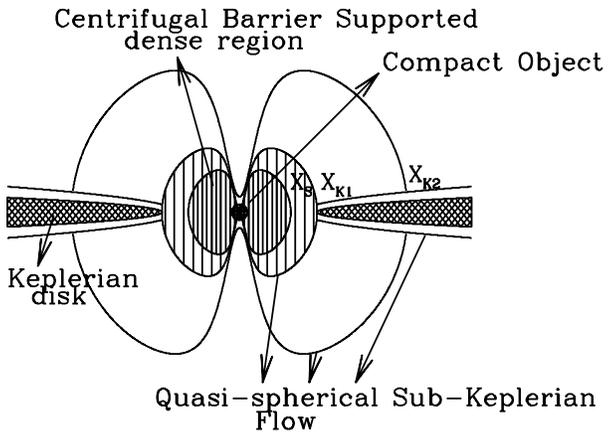
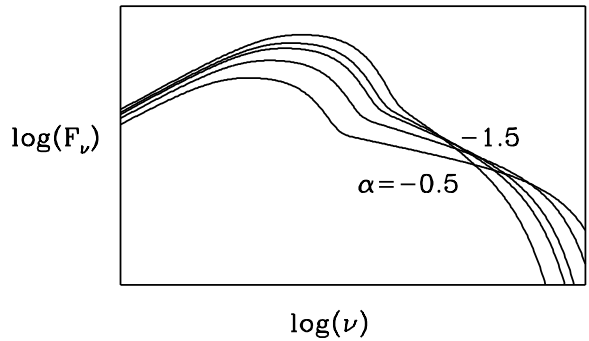
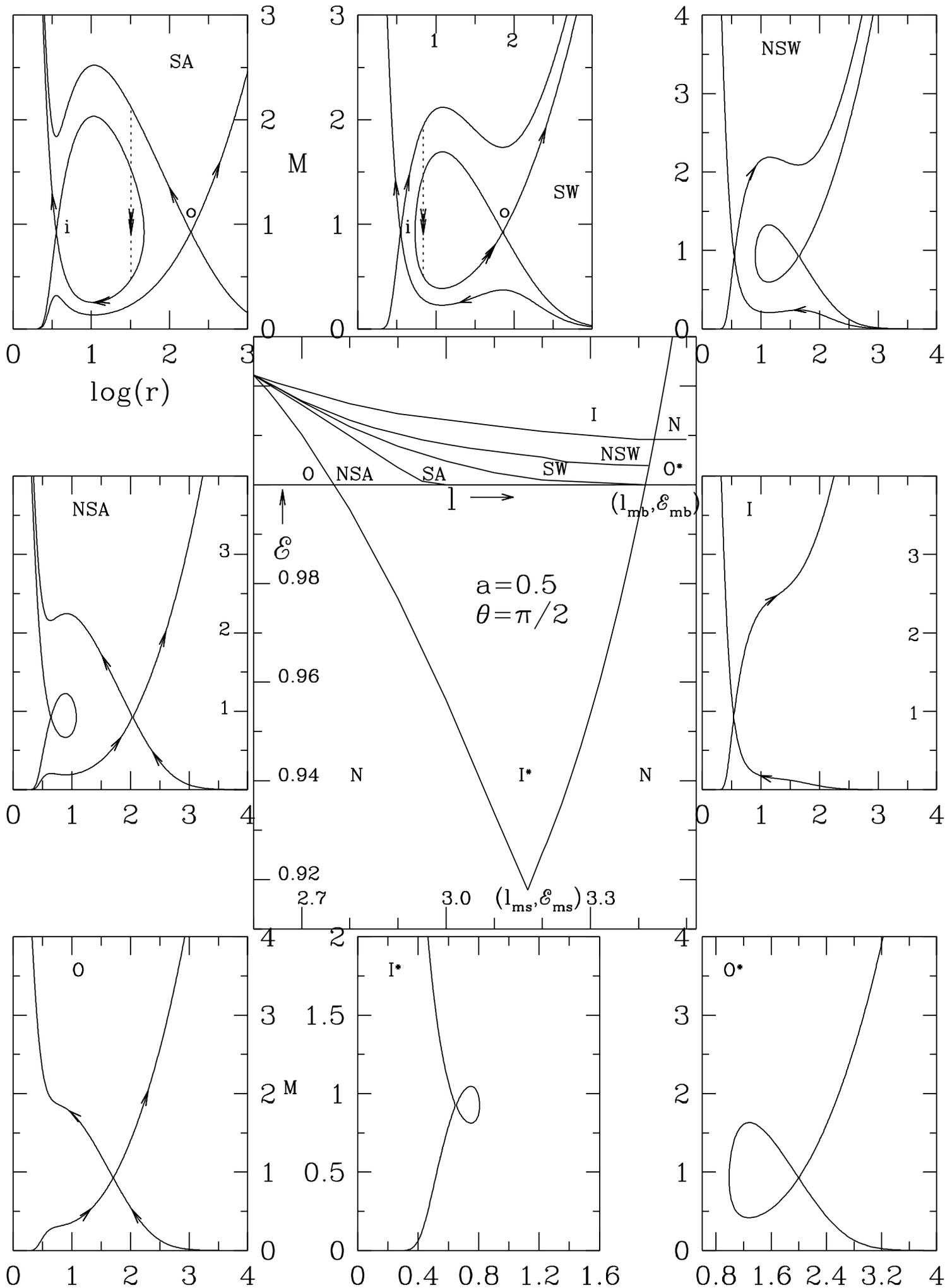
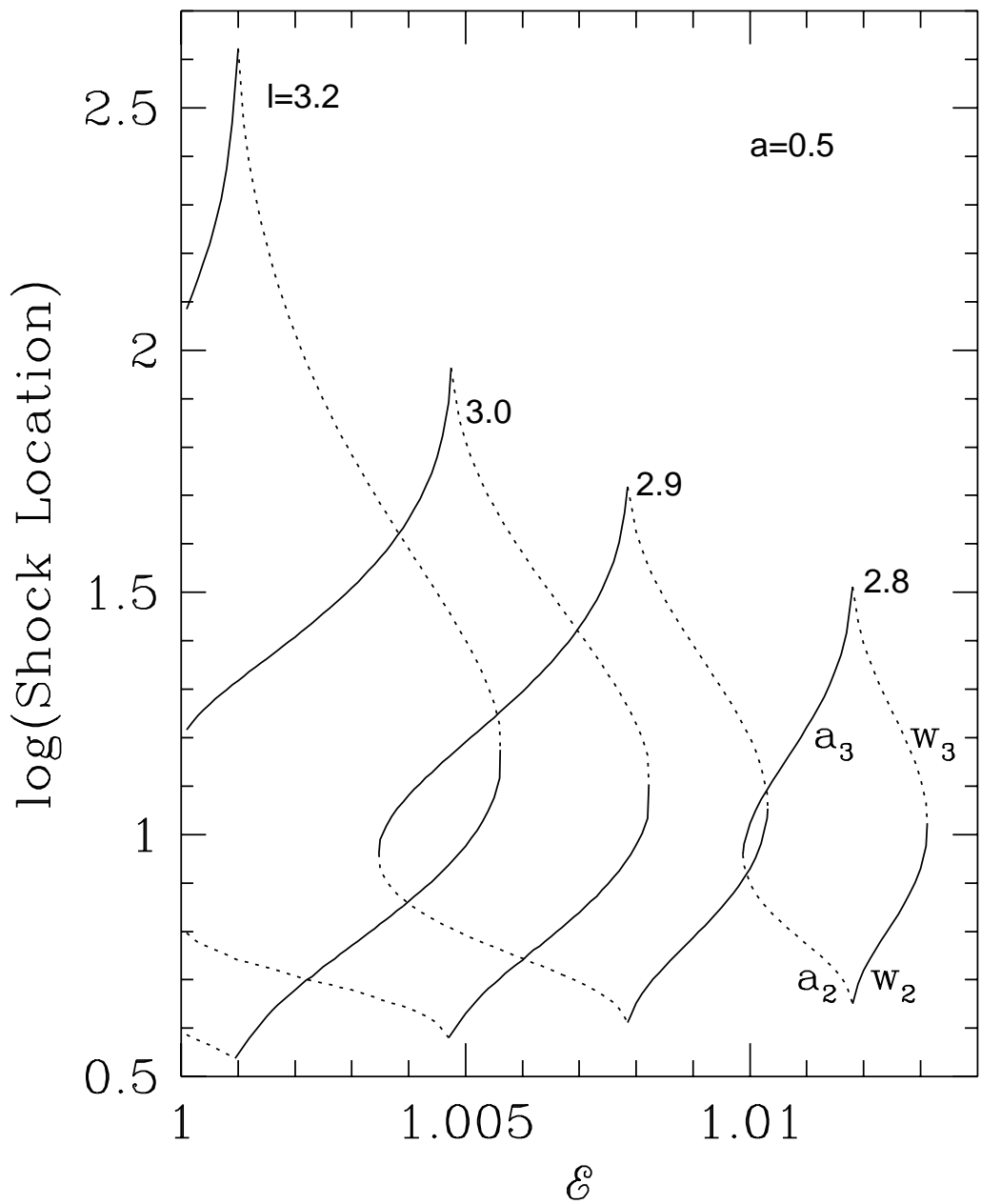
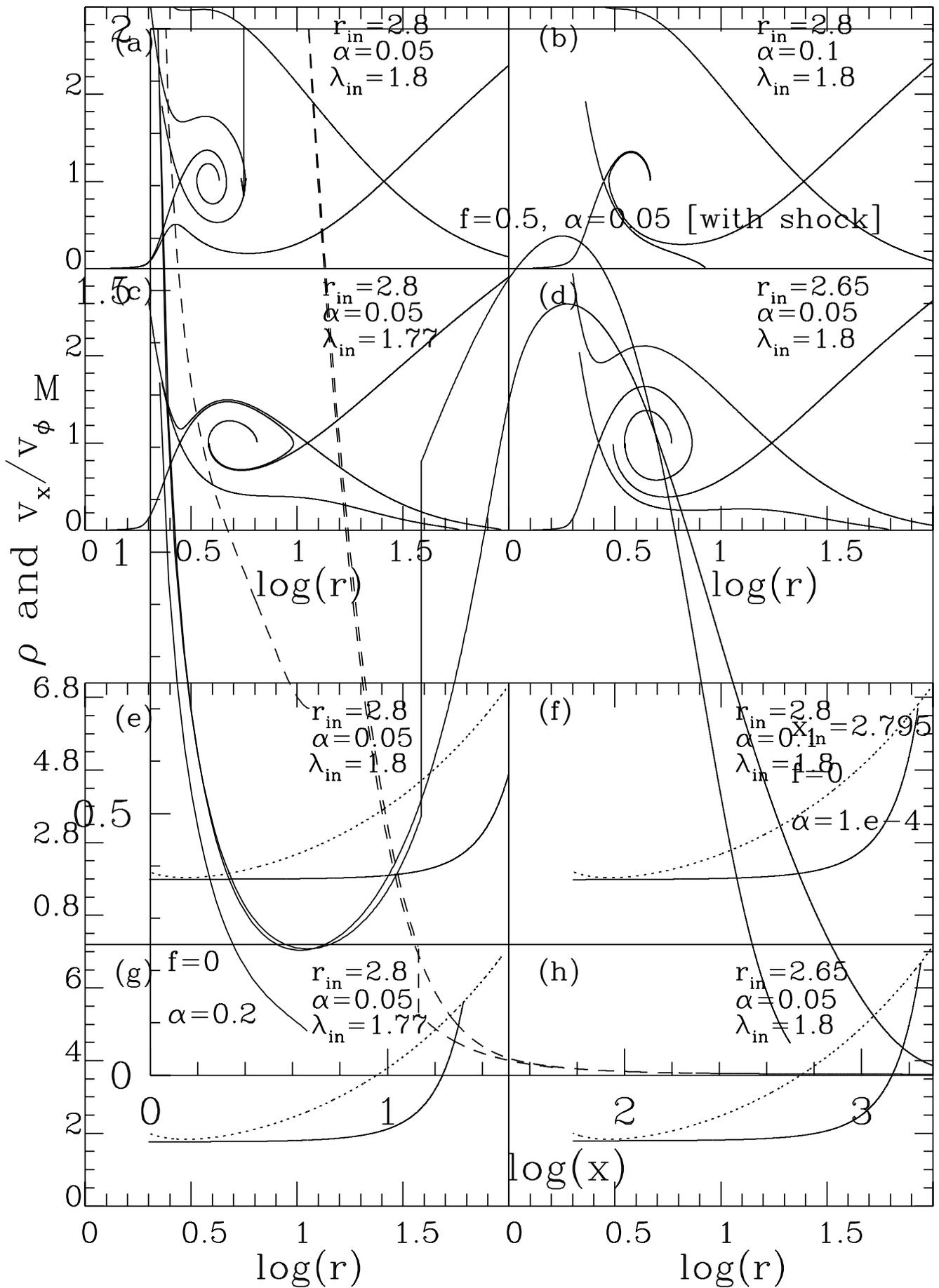
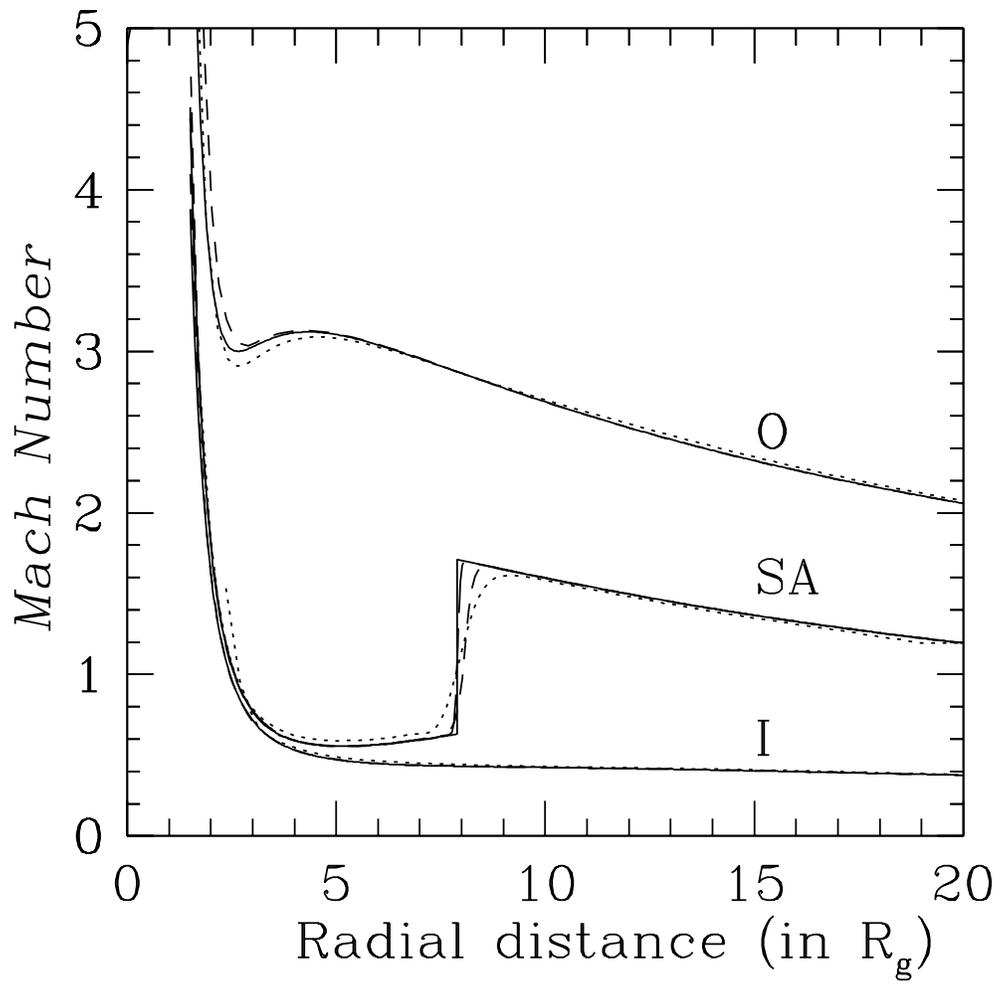
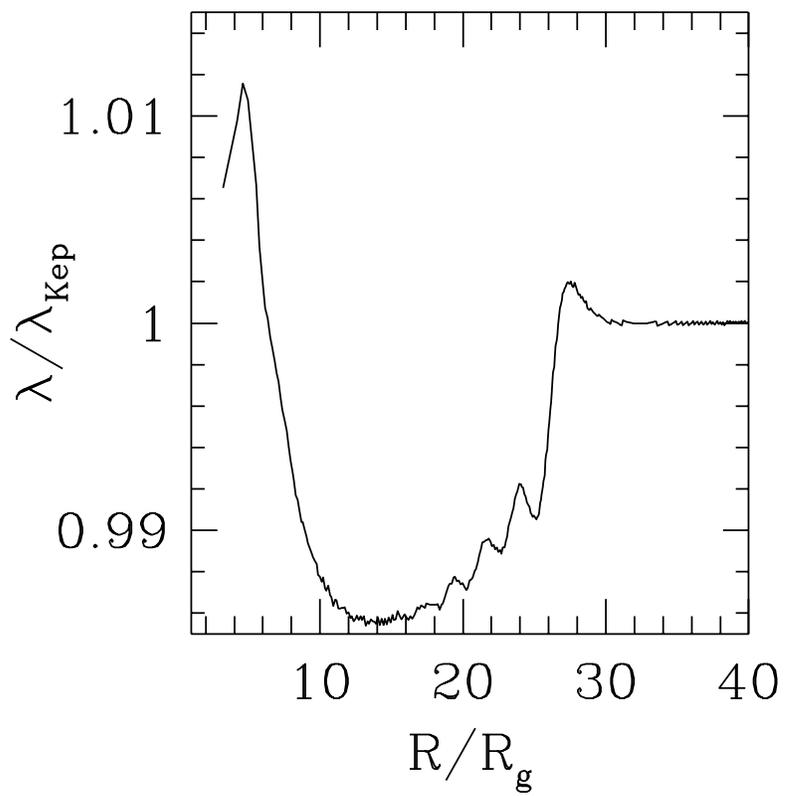
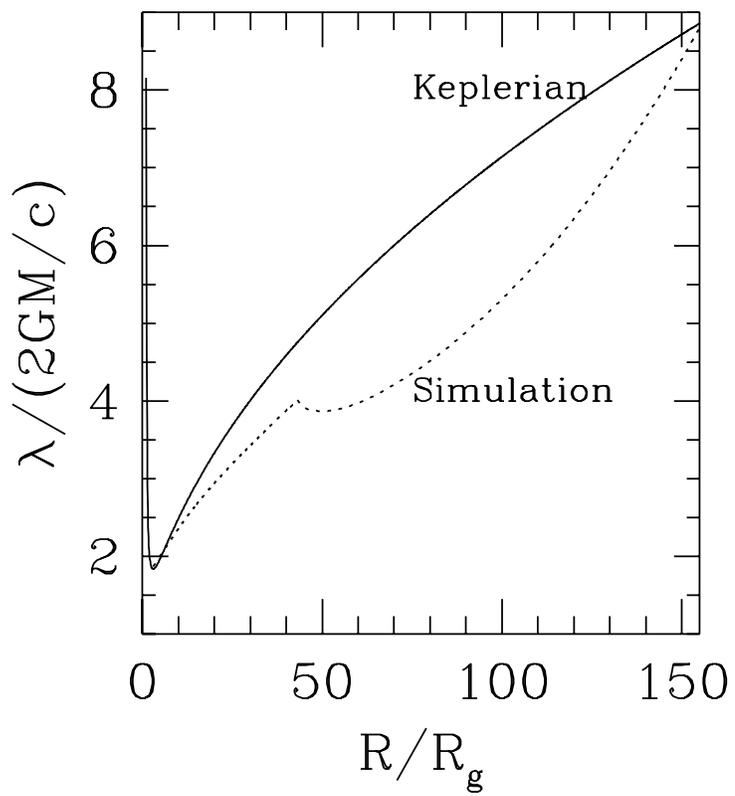
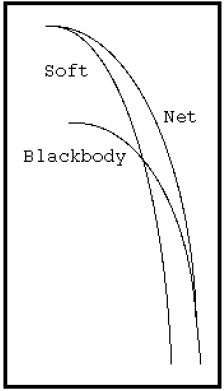
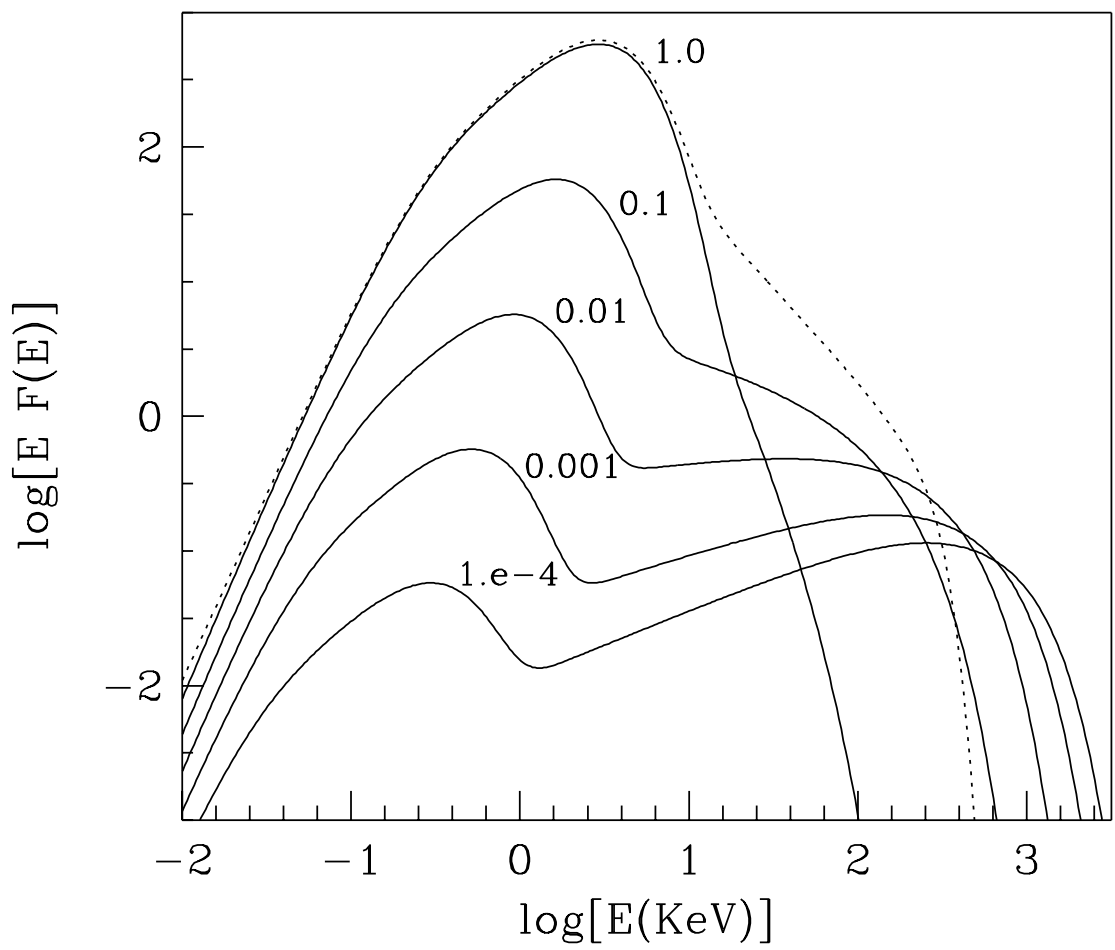
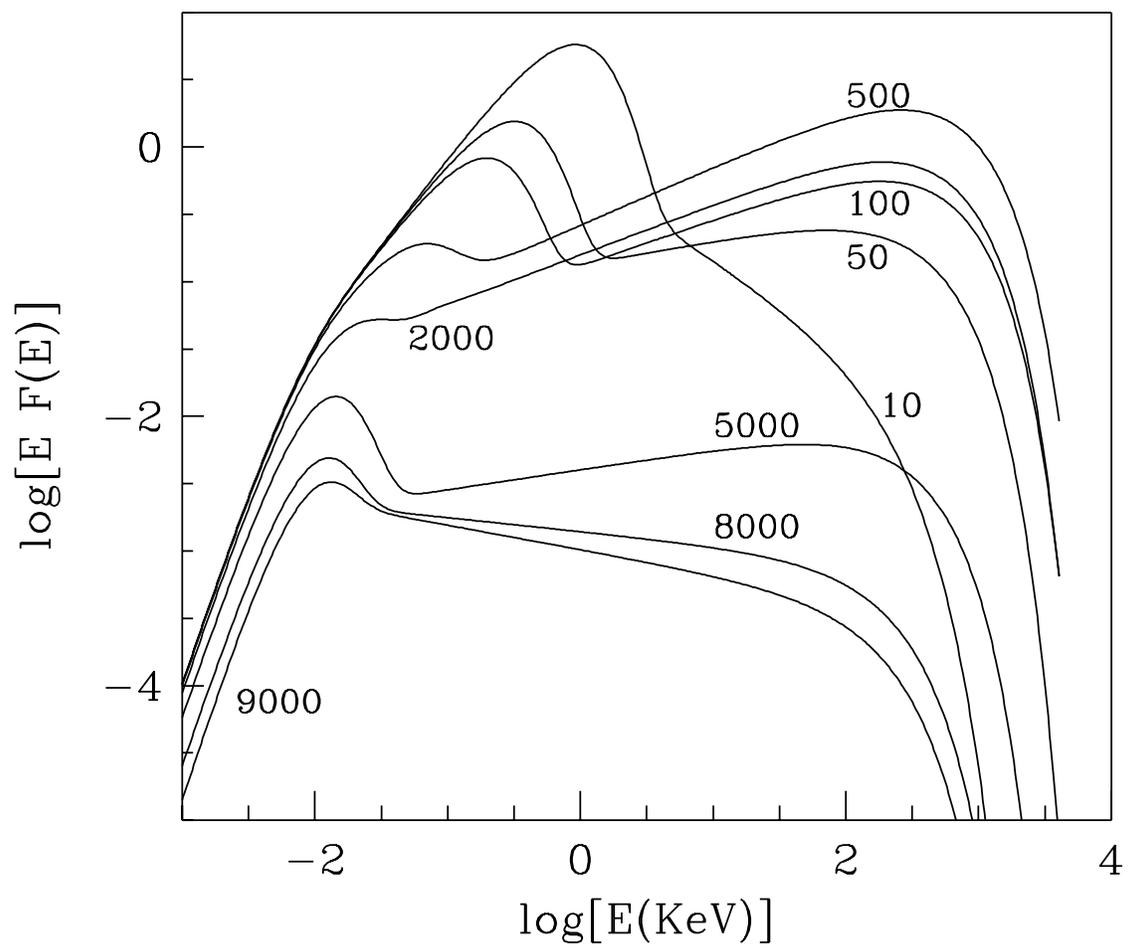
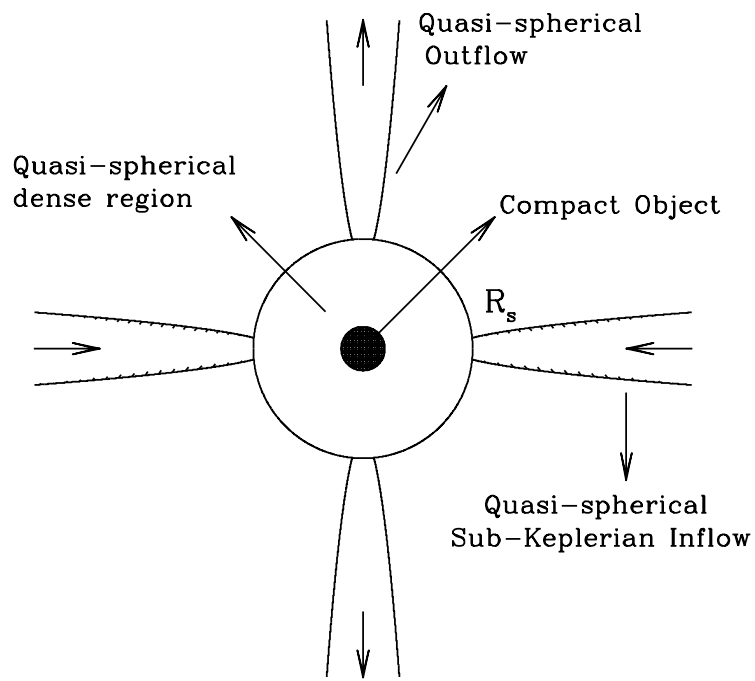
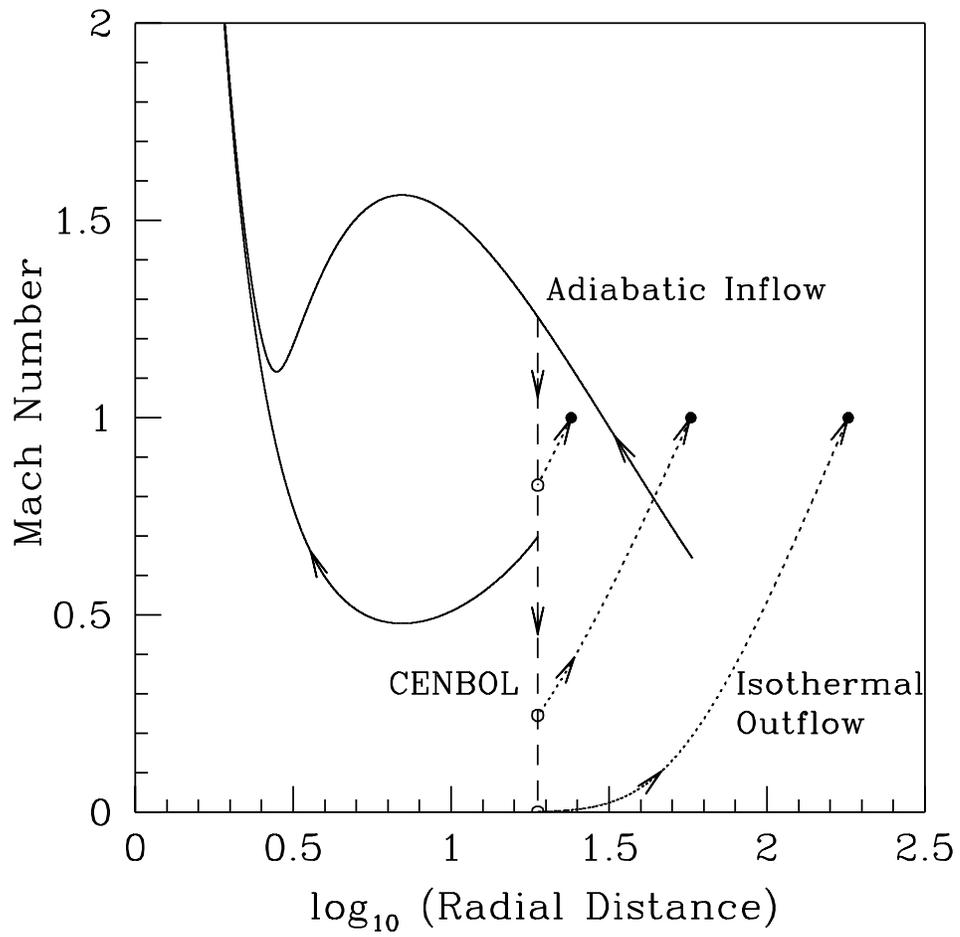
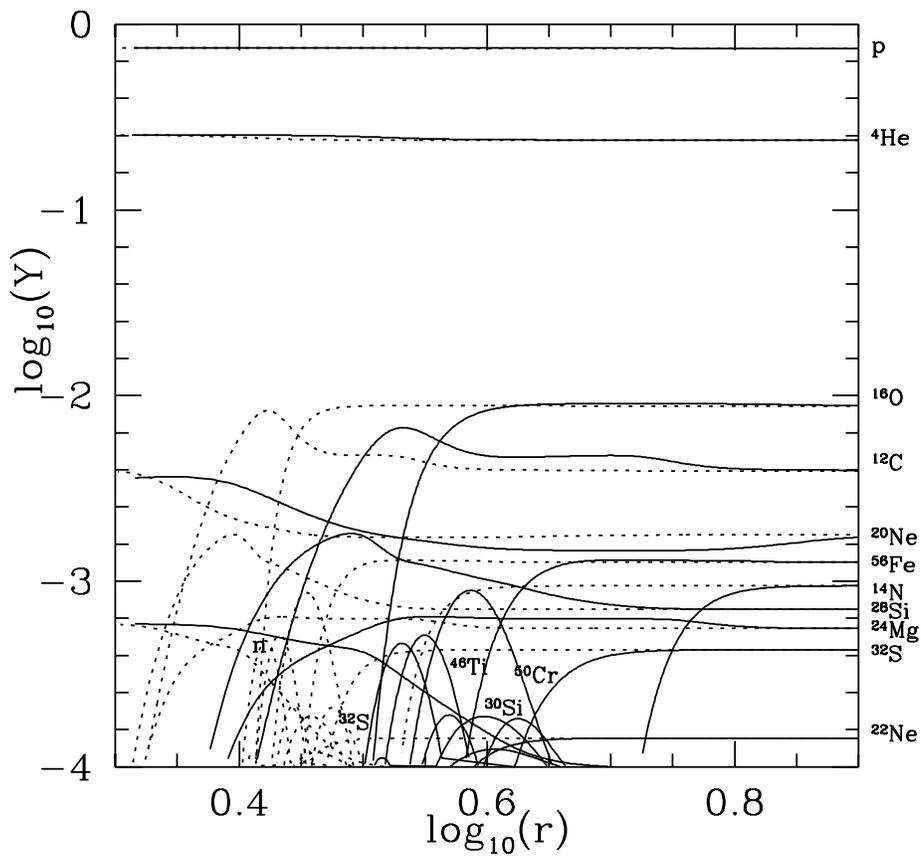
\end{document}